\documentclass[a4paper,fleqn]{cas-dc}

\usepackage[authoryear,longnamesfirst]{natbib}
\usepackage{enumitem}
\usepackage{float}
\usepackage{hyperref}
\usepackage{placeins}
\def\tsc#1{\csdef{#1}{\textsc{\lowercase{#1}}\xspace}}
\tsc{WGM}
\tsc{QE}

\begin{document}
\let\WriteBookmarks\relax
\def\floatpagepagefraction{1}
\def\textpagefraction{.001}

\shorttitle{Learning Team-Aware Intent in Multi-Agent Systems}    

\shortauthors{Nagarani et~al.}  

\title [mode = title]{MR-TGN: A Meta-Role Temporal Graph Network for Team-Level Intent Prediction in Multi-Agent Systems}  

\tnotemark[1] 

\tnotetext[1]{his work was supported by the Aeronautical Develop-
ment Agency under grant no. RB23241328AEADAX008980
and IIT Madras under grant no. RF24250403AENFIG008980.} 

\author[1]{Nagarani Brammanayagam}

\cormark[2]


\ead{nagarani.kavin@gmail.com}

\affiliation[1]{organization={Postdoctoral Researcher, Department of Aerospace Engineering},
            addressline={Indian Institute of Technology Madras}, 
            city={Chennai},
         citysep={},
            postcode={600036}, 
            state={Tamilnadu},
            country={India}}


\credit{Conceptualization of this study, Methodology, Software, Formal analysis, Writing – original draft.}

\author[2]{Devaprakash Muniraj}


\ead{deva@smail.iitm.ac.in}


\credit{Conceptualization of this study, Methodology, Supervision, Validation, Funding acquisition, Formal Analysis,  Writing – review \& editing.}

\affiliation[2]{organization={Assistant Professor, Department of Aerospace Engineering},
            addressline={Indian Institute of Technology Madras}, 
            city={Chennai},
         citysep={}, 
            postcode={600036}, 
            state={Tamilnadu},
            country={India}}

\cortext[2]{Corresponding author}



\begin{abstract}
Collective intent prediction in multi-agent systems focuses on predicting the shared objectives and future behaviours of groups of interacting agents. The problem is particularly challenging because collective intent emerges from complex interactions, evolving cooperation structures, and long-term behavioural dependencies among heterogeneous agents operating in dynamic and partially observable environments. Furthermore, functional roles adopted by agents are often latent, may change over time, and are rarely available as explicit annotations, making the learning of coordinated group behaviours significantly more difficult.
To address these challenges, this paper proposes a Meta-Role Temporal Graph Network (MR-TGN) framework for collective intent prediction in multi-agent systems. The proposed framework models agents as dynamically evolving graph entities and employs temporal memory mechanisms to encode historical interactions and coordination behaviours. To capture higher-level behavioural knowledge, MR-TGN introduces a memory-enhanced meta-role learning mechanism that derives latent role representations from agent-centric behavioral representations without requiring explicit role labels. In contrast to scenario-specific node memories, the learned role representations are stored as global behavioural knowledge that can be transferred across scenarios, thereby improving the representation of coordinated behaviours in previously unseen environments. By jointly modelling temporal interaction dynamics, evolving cooperation structures, and transferable role representations, MR-TGN enables the inference of high-level collective intents from coordinated multi-agent behaviours.
An evaluation methodology for early collective intent prediction is proposed to assess prediction accuracy and timeliness, enabling realistic evaluation of the model's ability to anticipate collective objectives during the early stages of mission execution. Experimental results on representative multi-agent scenarios demonstrate that the proposed framework consistently outperforms competitive baselines and achieves effective early prediction of collective intents in dynamic and adversarial environments.

\end{abstract}




\maketitle

\section{Introduction}\label{Intro}
Autonomous multi-agent systems comprise multiple interacting entities that operate in partially observable, uncertain, and continuously evolving environments. These entities perceive their surroundings, make decisions independently, and interact with other agents to achieve a collective objective. The behaviours of the agents are often interdependent because the actions of one agent can influence the decisions and responses of the others. Consequently, the overall system behaviour emerges from the interactions and coordination among multiple agents. Understanding these interactions is therefore essential for analysing collective behaviours and improving decision-making in autonomous multi-agent systems. In adversarial environments, these agents operate in the presence of opponents whose objectives conflict with their own, such environments are highly dynamic because agents continuously adapt their behaviours in response to changing mission objectives, environmental conditions, and the actions of the other agents. As a result, interaction patterns, cooperation structures, and functional roles may change over time as the agents form, dissolve, or reorganise into different groups to accomplish their goals. Effective decision-making in these settings requires the agents to continuously perceive, interpret, and anticipate the behaviours of other entities. The ability to understand emerging coordination patterns and predict future behaviours is therefore critical for proactive planning, adaptive decision-making, and timely responses to evolving~situations.

Opponent modelling addresses these challenges by seeking to infer the goals, strategies, and future actions of adversarial agents from their observed behaviours and interactions. It has been extensively studied in domains such as strategic games \cite{Liu2020MultiAgent, Ge2024ModelingRationality}, robotics \cite{Liu2023TrajectoryGames, Tao2025TrajectoryPrediction}, military \cite{Huo2025AutonomousAirCombat, Queralta2020CollaborativeMultiRobot} and autonomous systems \cite{Hou2022BehaviorReasoning, Busch2022A}, where understanding an opponent's decision-making process enables more effective coordination and strategic planning. Early approaches primarily relied on rule-based reasoning \cite{Herik2005rulebased}, probabilistic inference \cite{Wei2013SweetSpot}, and handcrafted behavioural representations \cite{Liu2023OpponentModeling}, often assuming static environments and predefined opponent models. However, modern multi-agent environments exhibit dynamic interactions and continuously evolving behaviours, making these assumptions insufficient. In such environments, accurately understanding opponents ultimately requires anticipating their high-level intentions, which frequently emerge from coordinated behaviours and evolving interactions among multiple agents rather than from the actions of individual entities in isolation. Therefore, effective opponent modelling requires learning representations that capture both temporal interaction dynamics and the underlying cooperation structures among the  agents.

Learning-based approaches have emerged as an effective alternative for modelling complex behaviours in autonomous multi-agent systems. Instead of relying on predefined rules or handcrafted behavioural representations, these approaches learn meaningful representations directly from observations of agent states, interactions, and trajectories. Recent studies have employed sequential models \cite{Du2025TrajectoryPrediction, Teng2021GRU}, graph-based methods \cite{Liu2020MultiAgent, Yu2026ShipTrajectory}, and deep learning techniques \cite{He2016OpponentModeling, Li2023VesselTrafficFlow} to capture spatial dependencies, temporal evolution, and relational information among multiple agents. By learning from historical observations, these methods can encode rich behavioural representations that reveal latent interaction patterns and provide a compact description of agent dynamics. Reinforcement learning methods \cite{Ma2021OpponentPortrait, Huo2025AutonomousAirCombat} focus on learning optimal decision-making policies by enabling agents to interact with the environment and maximize long-term rewards, thereby addressing how agents should act under varying situations. While these approaches have demonstrated remarkable success in autonomous control and cooperative decision making, they are not well suited to infer latent behavioural intent. 
In contrast, representation learning \cite{Zhang2020SpatioTemporalEEG, He2016OpponentModeling} approaches aim to learn informative latent representations that capture underlying behavioural patterns, providing descriptive features for various downstream tasks, including behaviour understanding, trajectory prediction, coordination analysis, and intent inference. Among these, collective intent prediction remains challenging because the intent of a team is often not explicitly observable from instantaneous states or short-term behaviours. Instead, it emerges through coordinated multi-agent interactions, requiring models that capture dynamic cooperation and long-term temporal dependencies. 

Although some recent representation learning methods \cite{Schneider2023LatentEmbeddings,Zhang2023BIPTree} have demonstrated the effectiveness of learning latent embeddings and prototype-based representations for behaviour modelling, they do not explicitly capture transferable functional roles that emerge from the behavioural characteristics of individual agents. Consequently, the learned representations are often tightly coupled to specific agents and scenarios, limiting their ability to generalize across different operational environments. Rather than relying on predefined role labels, it is therefore desirable to learn latent role behaviours from agent-centric behavioral representations, enabling transferable behavioural knowledge across diverse scenarios. 
Furthermore, multi-agent systems often comprise heterogeneous agents with diverse sensing capabilities, mobility characteristics, functional responsibilities, and decision-making strategies. Based on the mission and environment, agents may dynamically assume different functional roles, such as information providers, coordinators, decision-makers, or task~executors.  Although Temporal Graph Networks (TGNs) \cite{rossi2020temporalgraphnetworksdeep, Jeon2025Leveraging} implicitly encode these coordination patterns within spatio-temporal representations, the learned embeddings remain entangled and do not explicitly expose the latent functional roles of individual agents or their contribution to team-level coordination. Moreover, the node-centric memory mechanism in TGNs primarily preserves scenario-specific interaction histories, making it difficult to capture higher-level behavioral regularities required for reasoning about collective intent in heterogeneous teams.

Motivated by these challenges, this paper proposes a Meta-Role Temporal Graph Network (MR-TGN) framework for collective intent prediction in autonomous multi-agent systems. The proposed framework employs TGN to represent agents as dynamically evolving graph entities and continuously update node representations based on temporally ordered interactions. Building upon the rich spatio-temporal representations and node memories learned by TGN, MR-TGN introduces a memory-enhanced role-aware learning module that discovers latent functional roles from agent-centric behavioral dynamics without requiring explicit role annotations. To facilitate knowledge transfer across scenarios, the framework maintains a global latent role memory that captures transferable role-related behavioral knowledge and enables the reuse of learned role representations in previously unseen situations. By integrating scenario-specific interaction memories with a global role memory, MR-TGN learns representations that jointly capture temporal interaction dynamics, transferable functional roles, and coordinated team behaviours. These enriched representations provide the foundation for role assignment, team formation reasoning, and the inference of high-level collective intent in heterogeneous multi-agent systems.

\noindent{The main contributions of this work are summarised below:}
\begin{enumerate}[leftmargin=1.2em,labelsep=0.2em,itemsep=0.2em]
    \item We propose a novel \textit{MR-TGN} framework for team-level intent prediction in autonomous multi-agent systems. The proposed framework jointly leverages temporally evolving interactions, cooperation structures, and latent role representations to infer shared mission objectives emerging from coordinated behaviours in complex and adversarial environments.

    \item We introduce a role-aware representation learning mechanism that captures collective behaviours not only from spatial and temporal interactions among agents but also from the emergence of latent functional roles. The proposed approach learns implicit role representations directly from agent behavioural patterns without requiring explicit role annotations, enabling the modelling of evolving behaviours in heterogeneous multi-agent systems.

    \item We demonstrate the applicability of the proposed MR-TGN framework through an air combat intent prediction case study involving heterogeneous teams executing diverse mission objectives. To the best of our knowledge, this is the first application of role-aware temporal graph representation learning for collective intent prediction in autonomous air combat scenarios.

    \item We introduce an evaluation methodology for early collective intent prediction that assesses both prediction accuracy and prediction timeliness. The proposed evaluation provides a realistic measure of a model's ability to infer collective objectives at early stages of mission execution, thereby supporting proactive decision-making and timely responses in dynamic adversarial environments.
\end{enumerate}

 \section{Related Work}
 Opponent modelling has evolved considerably from early rule-based and probabilistic reasoning frameworks to modern data-driven learning approaches. Early research predominantly adopted rule-based reasoning frameworks \cite{Herik2005rulebased}, in which expert knowledge and manually designed rules were employed to describe opponent strategies and predict future actions. Subsequently, probabilistic approaches \cite{Wei2013SweetSpot} emerged to address uncertainty in opponent behaviour by estimating hidden intentions and maintaining belief distributions over possible actions. In parallel, handcrafted behavioural representations \cite{Liu2023OpponentModeling}, based on domain-specific features such as movement patterns, tactical formations, and predefined interaction templates, were developed to capture characteristic opponent behaviours. Such assumptions become increasingly restrictive in contemporary multi-agent systems, where interactions among agents are inherently dynamic, behaviours continuously evolve in response to changing objectives and environmental conditions, and coordination patterns may reorganise over time. 

\subsection{Learning-based Representation Models for Multi-Agent Reasoning}
Recent learning-based intent prediction approaches have increasingly exploited spatio-temporal representations to capture both spatial dependencies and temporal evolution in sequential data. 
The progression of learning-based approaches for multi-agent reasoning can be viewed as an evolution in the type and complexity of information encoded by the learned representations. To illustrate this progression, representative models are selected according to their distinct mechanisms for capturing spatial, temporal, and spatio-temporal dependencies which is summarized in Table~\ref{tab:model_capabilities}. Graph Attention Network (GAT) is selected as the representative spatial graph model because attention-based GNNs have become a fundamental paradigm for learning topology-aware node representations by adaptively weighting the importance of neighbouring nodes during message passing [\cite{Zhang2020A}, \cite{Sun2023AttentionSurvey}]. Among these models, GAT learns the relative importance of neighbouring agents through attention-based interaction aggregation, but processes each graph state independently without explicitly modelling temporal evolution. Long Short-Term Memory (LSTM) networks have been widely adopted for intent recognition because they effectively capture long-term temporal dependencies from sequential observations. For example, an LSTM-based framework combined with adversarial training was proposed to improve intention recognition from RGB video sequences by learning robust temporal representations [\cite{Mavsar2024GANLSTM}]. But they do not explicitly model relational interactions among multiple agents, limiting their applicability to coordinated multi-agent reasoning.

Beyond independent spatial and temporal modelling, spatio-temporal learning approaches jointly capture structural and temporal dependencies. For example, CNN-LSTM and GRU-CNN architectures learn spatial characteristics and temporal dynamics through cascaded convolutional and recurrent modules, demonstrating improved performance in trajectory prediction and intention recognition tasks \cite{Du2025CNNGRU, WANG2024CNNBiLSTM}. However, these methods primarily operate on structured sequential signals or individual trajectories and do not explicitly represent dynamically evolving inter-agent interactions. To address this limitation, graph-based spatio-temporal models have been proposed. Graph Neural Network GNN-LSTM and Spatio-Temporal Graph Convolution Network (STGCN) are included as representative graph-based spatial-temporal models.
Graph Neural Network–Long Short-Term Memory (GNN-LSTM) architectures represent a common approach for spatio-temporal learning by combining graph-based spatial modelling with recurrent temporal learning. 
For example, a graph-attention LSTM framework represented the traffic network as a graph, employed graph attention to capture spatial dependencies among connected nodes, and utilised LSTM to model temporal traffic dynamics [\cite{Zhang2022GraphAttentionLSTM, Vitulyova2025A}]. Such architectures factorize spatial interaction modelling and temporal sequence learning into successive stages, where graph representations are first extracted at each timestamp and subsequently processed by the recurrent network. Consequently, temporal reasoning operates on already aggregated spatial representations rather than directly modelling the evolution of pairwise interactions over time. Similarly, Spatio-Temporal Graph Convolutional Networks (STGCNs) jointly apply graph convolution and temporal convolution to learn spatial dependencies and temporal dynamics from graph-structured data [\cite{Luo2024STMGCN,Siabi2026Innovative}]. Their effectiveness has been demonstrated in applications such as terrorism risk forecasting, where multiple graph structures were combined with temporal modelling to capture complex spatio-temporal correlations \cite{Luo2024STMGCN}. Nevertheless, STGCNs primarily capture local spatio-temporal patterns, and their fixed temporal receptive field limits the modelling of long-range historical dependencies and irregularly evolving coordination behaviours.

\begin{table*}[t]
\centering
\caption{Comparison of representative models based on their spatial,
temporal, and spatio-temporal modelling capabilities.}
\label{tab:model_capabilities}

\small
\setlength{\tabcolsep}{2.5pt}
\renewcommand{\arraystretch}{1.25}

\begin{tabular}{p{3cm}ccccccccc}
\toprule

\textbf{Model}
& \makecell{\textbf{Spatial}\\\textbf{Relations}}
& \makecell{\textbf{Temporal}\\\textbf{Dependencies}}
& \makecell{\textbf{Spatial}\\\textbf{Attention}}
& \makecell{\textbf{Temporal}\\\textbf{Attention}}
& \makecell{\textbf{Local ST}\\\textbf{Patterns}}
& \makecell{\textbf{Long-Term}\\\textbf{History}}
& \makecell{\textbf{Persistent}\\\textbf{Memory}}
& \makecell{\textbf{Sequential}\\\textbf{Interaction}\\\textbf{History}}
& \makecell{\textbf{Transferable}\\\textbf{Latent Roles}} \\

\midrule

GAT [\cite{Zhang2020A, Sun2023AttentionSurvey}]
& $\checkmark$ & $\times$ & $\checkmark$ & $\times$
& $\times$ & $\times$ & $\times$ & $\times$ & $\times$ \\

Attention-LSTM [\cite{Mavsar2024GANLSTM}]
& $\times$ & $\checkmark$ & $\times$ & $\checkmark$
& $\times$ & $\checkmark$ & $\times$ & $\times$ & $\times$ \\

GNN-LSTM [\cite{Vitulyova2025A}]
& $\checkmark$ & $\checkmark$ & $\times$ & $\times$
& $\circ$ & $\checkmark$ & $\times$ & $\circ$ & $\times$ \\

Graph Attention LSTM [\cite{Zhang2022GraphAttentionLSTM}]
& $\checkmark$ & $\checkmark$ & $\checkmark$ & $\times$
& $\circ$ & $\checkmark$ & $\times$ & $\circ$ & $\times$ \\

STGCN [\cite{Luo2024STMGCN,Siabi2026Innovative}]
& $\checkmark$ & $\checkmark$ & $\times$ & $\times$
& $\checkmark$ & $\circ$ & $\times$ & $\circ$ & $\times$ \\

DySAT [\cite{Sankar2020DySAT:}]
& $\checkmark$ & $\checkmark$ & $\checkmark$ & $\checkmark$
& $\checkmark$ & $\checkmark$ & $\times$ & $\circ$ & $\times$ \\

TGN [\cite{rossi2020temporalgraphnetworksdeep, Jeon2025Leveraging}]
& $\checkmark$ & $\checkmark$ & $\checkmark$ & $\checkmark$
& $\checkmark$ & $\checkmark$ & $\checkmark$ & $\checkmark$ & $\times$ \\

MR-TGN (Proposed)
& $\checkmark$ & $\checkmark$ & $\checkmark$ & $\checkmark$
& $\checkmark$ & $\checkmark$ & $\checkmark$ & $\checkmark$ & $\checkmark$ \\

\bottomrule

\multicolumn{10}{l}{
\footnotesize
$\checkmark$: explicitly supported;
$\circ$: partially or indirectly supported;
$\times$: not explicitly supported.
}

\end{tabular}

\end{table*}

To address these limitations, recent studies have increasingly adopted dynamic graph-based learning frameworks, in which agents are represented as nodes and their interactions are modelled as time-evolving graphs, enabling the learning of relational dependencies that govern collective behaviors and intents. Among these approaches, Dynamic Self-Attention Network (DySAT) \cite{Sankar2020DySAT:}] and TGN [\cite{rossi2020temporalgraphnetworksdeep, Jeon2025Leveraging}]] are selected as representative dynamic graph representation learning models because they explicitly learn evolving spatio-temporal dependencies while maintaining expressive node representations for downstream reasoning tasks. Dynamic Self-Attention Network (DySAT) further introduces hierarchical attention by applying structural attention within individual graph snapshots and temporal self-attention across the sequence of node representations, enabling the model to identify both influential interactions and informative historical graph states. Nevertheless, temporal reasoning is performed over sequences of snapshot-level representations and does not maintain an explicit persistent state that is incrementally updated as new interactions are observed. TGN [[\cite{rossi2020temporalgraphnetworksdeep, Jeon2025Leveraging}]] addresses this limitation through node-wise memory that sequentially accumulates historical interaction information and combines the resulting memory states with temporal graph aggregation, making it suitable for modelling continuously evolving multi-agent interactions. However, the learned memory remains agent-specific and does not explicitly identify recurring functional behaviours or transfer shared role knowledge across agents and scenarios. Motivated by these complementary modelling capabilities and limitations, the proposed MR-TGN framework extends memory-based temporal graph learning with globally shared meta-role representations that capture recurring within-agent behavioural dynamics. By integrating evolving interaction information, persistent temporal memory, and transferable latent role knowledge, MR-TGN learns richer representations for team discovery and collective intent prediction across heterogeneous multi-agent systems.

Air combat provides a challenging case study for multi-agent intent prediction due to its highly dynamic, adversarial, and heterogeneous operational environment, where multiple agents with different capabilities and functional responsibilities interact and coordinate over time to accomplish mission objectives. Accurate prediction of opponent intent in such environments requires reasoning from the observed behaviours and interactions of multiple aircraft whose trajectories and coordination patterns continuously evolve during mission execution. Consequently, various studies have investigated air combat intent prediction using trajectory-based analysis, probabilistic reasoning, deep sequential models, and interaction-aware learning approaches. Deep learning-based air combat intent prediction has been explored using recurrent architectures such as GRU \cite{Teng2021GRU, xia2023incompletegru} and BiLSTM \cite{WANG2024CNNBiLSTM} to capture the temporal evolution of combat behaviours, while ANN-based frameworks \cite{wang2025attackintent} have also been developed for attack intent recognition in swarm engagement scenarios. To enhance temporal feature learning, hybrid 1D CNN–BiLSTM architectures \cite{Zhang20231DCNN-BiLSTM}  have been developed to jointly capture local temporal patterns and long-range sequential dependencies for air combat intention recognition. GNN-based relationship reasoning has been employed to capture complex inter-agent relationships for collaborative decision-making in large-scale air combat  \cite{haiyin2023TRIsonic}. Large-scale autonomous air combat has been modelled as a dynamic graph using multi-agent reinforcement learning, where automatic sub-team partitioning and message passing are employed to manage complex inter-agent relationships and support collaborative decision-making \cite{Fan2024subteam}. Collectively, these studies highlight the increasing use of deep learning and graph-based approaches for air combat intent recognition and collaborative decision-making.

\subsection{Collective Intent Prediction in Heterogeneous Multi-Agent Systems}

Although dynamic graph learning methods have significantly improved the modelling of evolving interactions, most existing intent prediction approaches infer intentions at the individual-agent level or directly classify mission intent from aggregated behavioural representations without explicitly reasoning about team formation. Consequently, they overlook the intermediate process through which coordinated groups emerge, limiting their ability to capture collective behaviours that arise from team organization and inter-agent cooperation.
In contrast, our framework explicitly performs team assignment prior to collective intent prediction, enabling intent reasoning to be grounded in dynamically inferred team structures rather than isolated agent behaviours.

Another important limitation is that existing studies rarely address collective intent reasoning in heterogeneous multi-agent teams, where agents possess distinct capabilities, responsibilities, and interaction behaviours. Existing studies on heterogeneous multi-agent systems have primarily focused on trajectory prediction by modelling how different categories of agents interact and influence one another. For instance, methods such as BIP-Tree employ interaction mechanisms that account for distinct behavioural perceptions and responses among heterogeneous agents and leverage high-order message passing to improve future trajectory forecasting \cite{Zhang2023BIPTree}. Despite these advances, the learned interactions and semantic representations are predominantly exploited to predict future motions rather than to reason about shared objectives emerging from coordinated behaviours. Consequently, the problem of inferring collective intent in heterogeneous teams remains largely underexplored.

To address these challenges, we propose a generic team-aware intent prediction framework based on MR-TGN and a sequential multi-task learning (MTL) paradigm. The MR-TGN encoder jointly models spatial coordination, temporal coordination, historical dependencies, and temporally synchronized interactions to learn coordination-aware representations of interacting agents. To further abstract these interaction patterns, the framework incorporates globally shared latent role memories that discover transferable functional behaviours directly from behavioral dynamics without requiring explicit role annotations. The resulting representations are subsequently exploited within a sequential multi-task decoder that first infers team structures and then predicts collective intent at the team level. Owing to its generic formulation, the proposed framework is applicable to a broad class of multi-agent systems involving coordinated behaviours and collective decision making. In this work, we employ air combat as a representative and challenging case study to demonstrate the effectiveness of the proposed~framework.

\section{Problem Formulation}
\label{sec:problem_formulation}

Consider a dynamic multi-agent environment consisting of a set of agents
$\mathcal{V}={v_1,v_2,\ldots,v_N}$ whose states and interactions evolve over time. The agents may exhibit heterogeneous motion characteristics, interaction patterns, and functional behaviours while coordinating to accomplish a common collective objective. Given the observed trajectories and interactions of the agents over a temporal observation window, the objective of this work is to identify the underlying team structure and predict the collective intent exhibited by the interacting agents.

\subsection{Dynamic Graph Representation}

The evolving multi-agent system is represented as a sequence of temporal interaction graphs, expressed as
\begin{equation}
\mathcal{G}^{1:T}
=
\left\{
\mathcal{G}^{1},
\mathcal{G}^{2},
\ldots,
\mathcal{G}^{T}
\right\}.
\end{equation}
where the interaction graph at time $t$ is defined as
\begin{equation}
\mathcal{G}^{t}
=
\left(
\mathcal{V}^{t},
\mathcal{E}^{t},
\mathbf{X}^{t},
\mathbf{F}^{t}
\right).
\end{equation}

Here, $\mathcal{V}^{t}$ and $\mathcal{E}^{t}$ denote the set of agents observed at time $t$ and the set of pairwise interactions among the observed agents, respectively. At each time step, the interaction graph is constructed as a fully connected graph, such that an edge $e_{ij}^{t}\in\mathcal{E}^{t}$ exists between every pair of distinct agents $v_i$ and $v_j$. Since the true coordination structure is unknown a priori, we construct a fully connected candidate interaction graph among all active agents to avoid prematurely discarding potentially meaningful interactions through heuristic edge pruning. The active node set changes with agent entry and departure, inducing corresponding topology changes. Each agent $v_i$ is associated with a node feature vector $\mathbf{x}_{i}^{t} \in \mathbb{R}^{d_n}$, which characterizes the instantaneous motion state and observable attributes of the agent, while each edge $e_{ij}^{t}$ is associated with an edge feature vector $\mathbf{f}_{ij}^{t}$ which represents the relative spatial and motion characteristics of agent $v_j$ with respect to agent $v_i$ at time $t$. As agents might enter or leave the observed environment over time, $\mathcal{V}^{t}$ and consequently, the corresponding edge set $\mathcal{E}^{t}$ might change. Thus, the graph topology dynamically evolves with the composition of the observed agent set, while maintaining full connectivity among all agents present at each time step. In addition, the time-varying node and edge features characterize the evolving states and pairwise relationships of the agents. Their evolution over time provides important information regarding changes in proximity, motion alignment, relative orientation, and coordinated movement patterns. The temporal observation history up to time $t$ is denoted~as $\mathcal{H}^{t} = \left\{
\mathcal{G}^{1}, \mathcal{G}^{2}, \ldots, \mathcal{G}^{t} \right\}$. Given $\mathcal{H}^{t}$, the proposed framework jointly addresses two related learning objectives: team assignment and collective intent prediction.


\section{Methodology}
\begin{figure*}[t]
    \centering
    \includegraphics[width=0.7\textwidth]{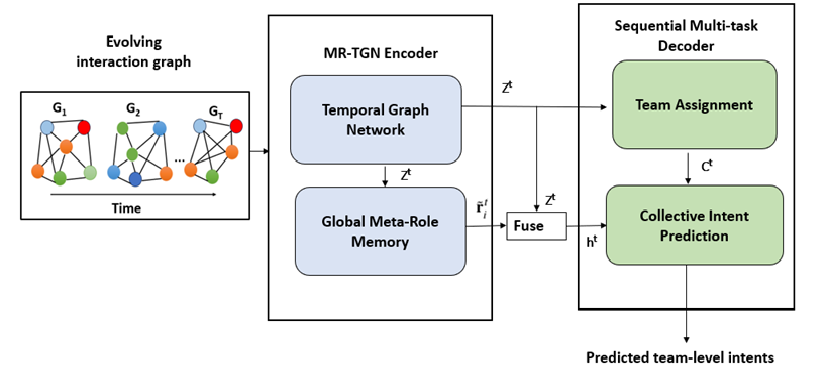}
    \caption{Overall architecture of the proposed framework.}
    \label{fig:architecture}
\end{figure*}

The overall architecture of the proposed framework is illustrated in Fig.~\ref{fig:architecture}. The input comprises multi-agent spatio-temporal trajectories containing kinematic and interaction features and relative motion information between interacting agents. These observations are represented as an evolving interaction graph, where nodes correspond to agents and edges encode their dynamic relationships over time. The proposed framework consists of two major components: (i) a MR-TGN encoder and (ii) a sequential MTL decoder. The MR-TGN encoder learns coordination-aware node representations by jointly modelling spatial coordination, temporal coordination, historical dependencies, and temporally synchronized/ sequenced interactions among agents. To further abstract these interaction patterns, the encoder incorporates a global latent role memory that discovers transferable functional roles from behavioural dynamics and enriches the node representations with role semantics. The sequential MTL decoder exploits these learned representations to reason about collective behaviour at multiple levels. First, coordination-aware node embeddings are utilized to infer team assignments by identifying groups of agents exhibiting coherent interaction patterns. Subsequently, the role-aware embeddings together with the inferred team structures are aggregated to predict a single collective intent for each team. By explicitly integrating coordination modelling, latent role discovery, and hierarchical reasoning over teams and intents, the proposed framework provides a unified approach for team-aware intent prediction in complex multi-agent environments.

\subsection{MR-TGN Encoder}
The proposed MR-TGN encoder is designed to learn agent representations that jointly capture the evolving coordination patterns and latent functional roles of agents in dynamic multi-agent environments. Given a sequence of time-evolving interaction graphs, the encoder first employs a TGN to model the spatial and temporal evolution of agent interactions. The resulting coordination-aware node representations are subsequently integrated with behavior-derived latent role representations, enabling the model to incorporate functional role information into the learned agent representations. Consequently, the MR-TGN encoder transforms the observed spatio-temporal interaction history, together with the inferred behavioral role information, into role-aware agent representations that provide a structured representation of the evolving multi-agent system for the downstream prediction tasks.
\begin{figure*}[t]
    \centering
    \includegraphics[width=0.8\textwidth]{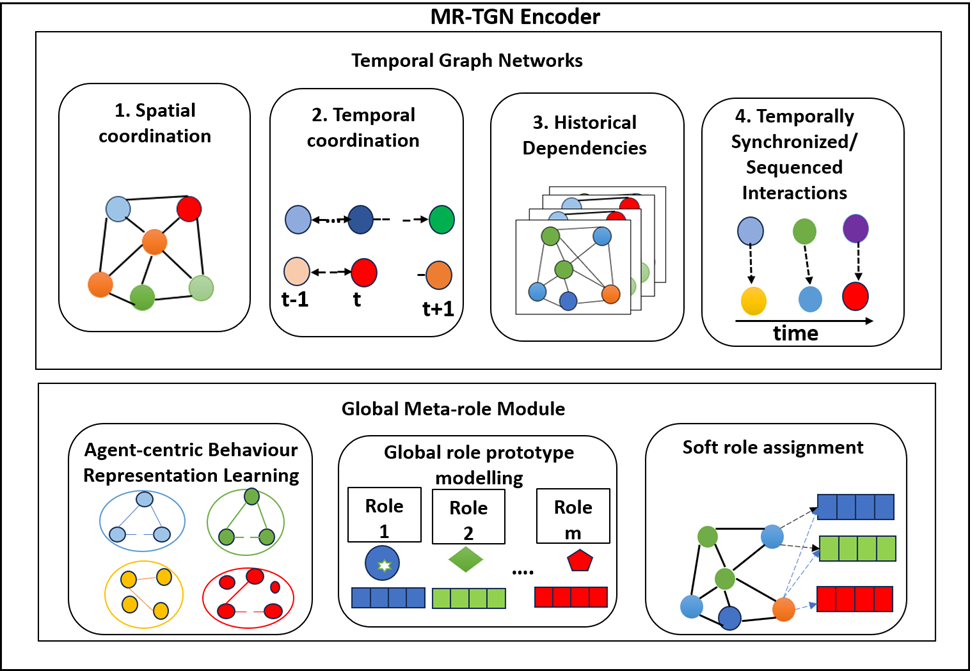}
    \caption{Overall architecture of the proposed MR-TGN encoder.}
    \label{fig:encoder}
\end{figure*}

\subsubsection{TGN for Coordination Representation Learning}
At each time step, the multi-agent system is represented as an evolving interaction graph. Since the interaction structure and motion states continuously evolve over time, analyzing individual graph snapshots independently is insufficient to capture the coordination dynamics underlying collective agent behavior. Therefore, the TGN component learns agent representations by integrating information from the current interaction structure with the temporally accumulated interaction history. As illustrated in Fig.~\ref{fig:encoder}, the TGN component is designed to capture four complementary aspects of coordination dynamics: spatial coordination, temporal coordination, historical dependencies, and temporally synchronized or sequenced interactions.

Spatial coordination characterizes the instantaneous relationships among agents at a particular time step. Through neighborhood message passing over the interaction graph, each agent aggregates information from its interacting neighbors and their pairwise edge features. This enables the encoder to represent local interaction structures such as relative proximity, motion compatibility, convergence or divergence, and other spatial dependencies that characterize coordinated behavior. Temporal coordination captures the evolution of agent states and interactions across consecutive time steps. Changes in motion states, neighborhood configurations, and pairwise interaction characteristics provide important evidence regarding how coordination patterns develop over time. By processing timestamped interactions and updating agent representations as new observations become available, the TGN models the temporal evolution of individual and collective behaviors.

Historical dependencies represent the influence of past interactions on the current behavior of an agent. In dynamic multi-agent environments, the current state of an agent may depend not only on its immediate interactions but also on previously observed coordination patterns. The memory mechanism of the TGN maintains and updates agent-specific historical information based on incoming temporal interactions, thereby enabling the current node representation to incorporate relevant information accumulated from earlier observations. Temporally synchronized or sequenced interactions capture dependencies arising from the relative timing and ordering of interactions among multiple agents. Coordinated behavior may emerge when different agents perform related actions simultaneously or according to a particular temporal sequence. Such dependencies cannot be fully represented by considering only individual temporal trajectories or isolated graph snapshots. By processing temporally ordered interactions and integrating them with the historical memory states of interacting agents, the TGN learns representations that reflect both synchronized and sequential patterns of multi-agent coordination. The learned representation $\mathbf{z}_i^t$ summarizes the current motion state of agent $i$, its interactions with neighboring agents, relevant historical information, and the temporal evolution of its coordination relationships.

The TGN module processes time-stamped agent interactions in temporal mini-batches. For each batch associated with time $t$, the module follows a strictly causal workflow: (i) update node memories using messages generated in previous batches, (ii) compute temporal node embeddings by jointly leveraging the updated node memories together with the current batch’s node features and edge attributes, and (iii) generate raw messages from the current batch interactions for use in subsequent batches.

\subsubsection*{Memory update using past interactions}

Before processing interactions at time $t$, each node retrieves its historical interaction information from the Raw Message Store (RMS). The RMS stores raw messages generated by interactions at earlier timestamps that have not yet been incorporated into the node memory. Let $m_{u \rightarrow i}(\tau)$ denote a raw message generated by an interaction between nodes $u$ and $i$ at timestamp $\tau$. During batch processing, a node may receive multiple messages originating from different interactions. To handle this, all raw messages generated for node $i$ at timestamps strictly earlier than the current batch time $t$ are aggregated as
\begin{equation}
\bar{m}_i(t) = \mathrm{AGG}\left\{ m_{u \rightarrow i}(\tau) \;\middle|\; \tau < t \right\},
\label{eq:message_aggregation}
\end{equation}
where $i$ and $u$ denote node indices, $\tau$ represents the timestamp at which the interaction occurred, and $\mathrm{AGG}(\cdot)$ is a permutation-invariant batch-level aggregation operator. Herein, $\mathrm{AGG}(\cdot)$ is instantiated as the mean operator that summarizes the set of historical messages by averaging all messages received by $i$ prior to time~$t$.

After aggregating past interaction messages, the memory of node $i$ is updated using a gated recurrent unit (GRU) as 
\begin{equation}
s_i(t) = \mathrm{GRU}\!\left(\bar{m}_i(t),\, s_i(t^-)\right),
\label{eq:memory_update}
\end{equation}
where $s_i(t^-)$ denotes the pre-interaction memory state, and $\bar{m}_i(t)$ represents the aggregated messages derived from past interactions. This update is performed only for nodes that participate in at least one interaction event within the batch; the memory states of all other nodes remain unchanged. The GRU gating mechanism enables selective retention and forgetting of past information, allowing the memory to capture long-term interaction dependencies while remaining responsive to recent behavioral patterns. This causality-preserving memory update forms the foundation for subsequent node embedding computation and attention-based neighbor aggregation in the proposed framework.
\vspace{3mm}

\subsubsection*{Multi-head temporal attention}

After updating the node memories, the encoder computes a context-aware temporal representation for each agent. This representation integrates three complementary sources of information: (i) the historical interaction context retained in the node memory, (ii) the agent's current behavioral state, and (iii) the current pairwise relationships with neighboring agents. The first two sources are combined to form the node representation, while the relational information is incorporated during the subsequent attention computation.
Specifically, the context representation of node $i$ is defined as
\begin{equation}
\mathbf{c}_i^{t}
=
\left[
\mathbf{s}_i^{t^-}
\,\Vert\,
\mathbf{x}_i^{t}
\right]
\label{eq:context_vector}
\end{equation}
where $\mathbf{s}_i^{t^-}$ denotes the memory state of node $i$
immediately before time $t$, and $\mathbf{x}_i^{t}$ represents its current node features. To incorporate the heterogeneous relational information between agents, the edge feature vector $\mathbf{e}_{ij}^{t}$ is transformed into a learned edge representation as
\begin{equation}
\mathbf{g}_{ij}^{t}
=
f_e\!\left(\mathbf{e}_{ij}^{t}\right)
\label{eq:edge_encoding}
\end{equation}
where $f_e(\cdot)$ denotes a learnable edge-encoding function.

The query representation is derived only from the current agent because it reflects the information required to update its representation. In contrast, the key and value representations additionally incorporate the pairwise edge features, allowing the attention mechanism to evaluate neighboring agents based on both their individual states and their current relationships with the current agent. Accordingly, for each attention head $k$, the query ($\mathbf{q}_i^{(k)}$), key ($\mathbf{k}_{ij}^{(k)}$), and value ($\mathbf{v}_{ij}^{(k)}$) representations are computed as
\begin{multline}
\mathbf{q}_i^{(k)}
=
\mathbf{W}_Q^{(k)}
\mathbf{c}_i^{t}
\quad \text{(Query)}, \\
\mathbf{k}_{ij}^{(k)}
=
\mathbf{W}_K^{(k)}
\left[
\mathbf{c}_j^{t}
\,\Vert\,
\mathbf{g}_{ij}^{t}
\right]
\quad \text{(Key)}, \\
\mathbf{v}_{ij}^{(k)}
=
\mathbf{W}_V^{(k)}
\left[
\mathbf{c}_j^{t}
\,\Vert\,
\mathbf{g}_{ij}^{t}
\right]
\quad \text{(Value)}.
\label{eq:key_value}
\end{multline}

where $\mathbf{W}_{Q}^{(k)}$, $\mathbf{W}_{K}^{(k)}$, and
$\mathbf{W}_{V}^{(k)}$
are the learnable projection matrices for the $k$-th attention head that
transform the node context vector and the concatenated node-edge representation
into the query, key, and value spaces, respectively. For each attention head, the encoder estimates the relevance of every neighboring agent to the current agent by comparing the query with the edge-aware key representations of its neighbors. The resulting attention coefficients determine how strongly information from each neighboring agent contributes to the updated temporal representation. By conditioning the key and value representations on the learned edge embeddings, the attention mechanism learns coordination patterns that emerge from spatial relationships, temporal evolution, historical memory, and synchronized interactions among heterogeneous agents. Consequently, agents that exhibit stronger coordination receive higher attention weights regardless of their individual roles or feature similarity.

The unnormalized attention score between node $i$ and neighbor $j$ for 
attention head $k$ is then computed using scaled dot-product attention as
\begin{equation}
\psi_{ij}^{(k)}(t)
=
\frac{
\left(\mathbf{q}_i^{(k)}(t)\right)^{\top}
\mathbf{k}_{ij}^{(k)}(t)
}{
\sqrt{d_k}
},
\label{eq:attention_score}
\end{equation}

\noindent where $d_k$ denotes the dimensionality of the query and key 
representations. 
Unlike an attention mechanism based on predefined trajectory similarity 
or pairwise interaction recency, the proposed formulation learns the 
relative importance of neighboring agents directly from their historical 
memory states, current behavioral features, and pairwise relational 
context. This formulation is particularly suitable for heterogeneous 
multi-agent teams, where coordinated agents may exhibit dissimilar motion 
patterns due to their distinct functional behaviors. Conditioning the 
attention mechanism on the complete edge representation allows MR-TGN to 
identify informative interactions without imposing an explicit assumption 
that behaviorally similar agents should receive higher attention weights.
 The normalized attention weights are computed as 
\begin{equation}
\alpha_{ij}^{(k)}(t)
=
\frac{
\exp\!\big(\psi_{ij}^{(k)}(t)\big)
}{
\sum\limits_{u \in \mathcal{N}(i,t)}
\exp\!\big(\psi_{iu}^{(k)}(t)\big)
}\,.
\label{eq:attention_weight}
\end{equation}

\noindent Each attention head computes a self-aware temporal embedding by combining a coordination-aware aggregation of neighboring nodes with a query derived from the node’s own memory and current state features as given by
\begin{equation}
\mathbf{z}^{\mathrm{t}}_{i,(k)}
=
\sum_{j \in \mathcal{N}(i,t)}
\alpha^{(k)}_{ij}(t)\,\mathbf{v}^{(k)}_{ij}(t)
+
\mathbf{q}^{(k)}_{i}(t).
\label{eq:head_aggregation}
\end{equation}

\noindent The final temporal embedding is given by the concatenation of all heads as 
\begin{equation}
\mathbf{z}_i^{\mathrm{t}}
=
\big\Vert_{k=1}^{K}
\mathbf{z}_{i,(k)}^{\mathrm{t}}.
\label{eq:multihead}
\end{equation}

\subsubsection*{Raw message computation for future memory updates}

\noindent After computing $z_i^{\mathrm{t}}$, the model encodes the interaction 
observed at time $t$ into a raw message that will influence future timestamps. The message for interaction $(i,j,t)$ is computed as  
\begin{equation}
m_{i\to j}(t)
=
\mathrm{MLP}_m\big(
z_i(t^-),\,
z_j(t^-),\,
e_{ij}(t)
\big) 
\label{eq:message_function}
\end{equation}
\noindent This message is stored in the RMS and is used when updating the memory for future interactions. No memory update is performed at time $t$, thereby preserving strict temporal causality.

\subsubsection{Global Meta-Role Module}
Although the TGN captures how agents interact and how coordination relationships evolve over time, agents exhibiting similar behavioural dynamics may perform similar functional roles within the multi-agent system. Such roles are generally not directly observable and may recur across different teams, scenarios, and interaction configurations. For example, in an air combat scenario, agents exhibiting similar behavioural dynamics may consistently assume functional roles such as escort, penetrator, flanker, decoy, or reconnaissance, regardless of their identities or the specific mission. Although these roles are not explicitly provided during training, they can emerge naturally from the learned behavioural representations. To capture this higher-level behavioural regularity, the proposed encoder introduces a behaviour-guided meta-role module that first constructs an agent-centric behavioural reference from recent temporal dynamics and subsequently uses it to extract behaviour-relevant information from the context-rich TGN embeddings for latent role discovery. The meta-role module consists of three conceptual stages: agent-centric behaviour representation learning, global role prototype modelling, and soft role assignment. The resulting role representations are optimized jointly with the downstream team assignment and intent prediction objectives, enabling the global role memory to capture transferable latent behavioral patterns that support collective intent reasoning across different scenarios.

\subsubsection*{Agent-centric Behaviour Representation Learning}

The TGN encoder produces context-rich node embeddings $\mathbf{z}_{i}^{t}$ that jointly capture individual motion characteristics, inter-agent interactions. While these representations are effective for modeling collective behavior, directly using the complete node embeddings for role discovery may cause the learned role representations to be influenced predominantly by interaction or team-specific coordination patterns. In contrast, latent roles are expected to reflect recurring behavioral characteristics of individual agents while retaining the task-relevant contextual information learned by the temporal graph encoder.

To provide an explicit agent-centric behavioral reference, a behavioral representation is constructed from the recent temporal evolution of the node features. Specifically, let $\mathbf{X}_{i,\mathrm{beh}}^{t}
= \left[ \mathbf{x}_{i,\mathrm{beh}}^{t-L}, \ldots,
\mathbf{x}_{i,\mathrm{beh}}^{t} \right]$.
The representation $\mathbf{X}_{i,\mathrm{beh}}^{t}$ denotes a short temporal window of behavior-related node features for agent $i$, where $L$ determines the behavioral observation horizon. The behavioral feature vector $\mathbf{x}_{i,\mathrm{beh}}^{t}$ contains attributes that characterize the agent's intrinsic state evolution and individual behavioral dynamics, while excluding explicit relational information derived from neighboring agents or pairwise interactions. Depending on the application domain, these features may represent motion characteristics, state transitions, action patterns, or other agent-specific temporal attributes. The specific behavioral node features used in the air combat case study are described in Section ~\ref{sec:case_study}.

Rather than directly using these node-feature dynamics for role retrieval, we first summarize the recent behavioral dynamics into a compact behavioral descriptor. Specifically, temporal variations in the behavior-related node features are aggregated through mean pooling to capture the agent's recent behavioral evolution as $\mathbf{d}_{i,\mathrm{beh}}^{t}
=
\operatorname{Pool}
\left(
\Delta \mathbf{X}_{i,\mathrm{beh}}^{t}
\right)$,
where $\operatorname{Pool}(\cdot)$ denotes mean pooling over the temporal dimension. The resulting behavioral descriptor is then mapped by a lightweight projection network $\phi_{\mathrm{beh}}(\cdot)$ into a latent behavioral reference space as $\mathbf{h}_{i,\mathrm{beh}}^{t}
=
\phi_{\mathrm{beh}}
\left(
\mathbf{d}_{i,\mathrm{beh}}^{t}
\right)$. 
The resulting behavioral reference representation provides an agent-centric description of recent behavioral dynamics and is used to guide the extraction of behavior-relevant information from the context-rich TGN embedding.

Accordingly, a behavior-conditioned projection is introduced as $\mathbf{g}_{i}^{t} =
\sigma
\left(
\mathbf{W}_{b}\mathbf{h}_{i,\mathrm{beh}}^{t}
+
\mathbf{b}_{b}
\right)$, 
followed by
\begin{equation}
\mathbf{b}_{i}^{t}
=
\mathbf{z}_{i}^{t}
\odot
\mathbf{g}_{i}^{t}
\label{eq:behavior_conditioned_embedding}
\end{equation}
where $\sigma(\cdot)$ denotes the sigmoid activation function and the operator $\odot$ represents element-wise multiplication. $\mathbf{g}_{i}^{t}$ acts as a behavior-conditioned feature gate that selectively emphasizes the dimensions of the context-rich temporal embedding $\mathbf{z}_{i}^{t}$ that are relevant to the agent's recent behavioral dynamics. Consequently, $\mathbf{b}_{i}^{t}$ biases the context-rich TGN embedding toward agent-centric behavioral characteristics relevant to role discovery.

\subsubsection*{Global role prototype modelling}
Next, the behaviour guided representations in \eqref{eq:behavior_conditioned_embedding} are compared with a set of learnable global role prototypes to discover latent functional roles. Agents exhibiting similar behavioural patterns occupy nearby regions of the learned representation space. Consequently, recurring patterns in the representations can reveal underlying functional similarities among agents without requiring explicit role annotations. To represent these recurring behavioral patterns, the proposed framework maintains a set of $K$ learnable global role prototypes, given by
\begin{equation}
\mathcal{R}
=
\left\{
\mathbf{r}_{1},
\mathbf{r}_{2},
\ldots,
\mathbf{r}_{K}
\right\},
\qquad
\mathbf{r}_{k}
\in
\mathbb{R}^{d_r}
\label{eq:global_role_prototypes}
\end{equation}
where each prototype represents a latent functional role that captures recurring behavioural patterns shared across agents and training scenarios. The role prototypes are model parameters learned jointly with the remaining network components through end-to-end optimization. Unlike scenario-specific role representations, the global prototypes provide a shared latent role space in which recurring behavioural patterns can be captured across different scenario configurations and team compositions.

\subsubsection*{Soft role assignment}
For each agent, the behavior-conditioned representation is compared with the global role prototypes to determine its association with the latent roles. Rather than assigning an agent to a single role using a discrete decision, the proposed method employs a soft matching mechanism. The role-assignment probability of agent $v_i$ to the $k^{th}$ role prototype is expressed as
\begin{equation}
\alpha_{ik}^{t}
=
\frac{
\exp\left(
\operatorname{sim}
\left(
\mathbf{b}_{i}^{t},
\mathbf{r}_{k}
\right)
/\tau
\right)
}{
\displaystyle
\sum_{l=1}^{K}
\exp\left(
\operatorname{sim}
\left(
\mathbf{b}_{i}^{t},
\mathbf{r}_{l}
\right)
/\tau
\right)
}
\label{eq:role_assignment}
\end{equation}
where $\operatorname{sim}(\cdot,\cdot)$ denotes the similarity function and $\tau$ is a temperature parameter controlling the sharpness of the role assignments. The resulting role-assignment vector
\begin{equation}
\boldsymbol{\alpha}_{i}^{t}
=
\left[
\alpha_{i1}^{t},
\ldots,
\alpha_{iK}^{t}
\right]
\label{eq:role_assignment_vector}
\end{equation}
represents the degree to which the behavioral dynamics of an agent is associated with each latent role.

The role representation of an agent is subsequently obtained as the weighted combination of the global role prototypes as 
\begin{equation}
\tilde{\mathbf{r}}_{i}^{t}
=
\sum_{k=1}^{K}
\alpha_{ik}^{t}\mathbf{r}_{k}
\label{eq:soft_role_representation}
\end{equation}
This soft role representation allows an agent to exhibit characteristics associated with multiple latent roles and provides a differentiable mechanism for jointly learning the global role prototypes and the remaining model components.

Finally, the coordination-aware temporal representation and the inferred role representation are integrated to obtain the role-enhanced agent representation:
\begin{equation}
\mathbf{h}_{i}^{t}
=
\operatorname{Fuse}
\left(
\mathbf{z}_{i}^{t},
\tilde{\mathbf{r}}_{i}^{t}
\right),
\label{eq:role_enhanced_representation}
\end{equation}
where $\operatorname{Fuse}(\cdot,\cdot)$ denotes the feature fusion operation employed by the proposed framework. The resulting representation preserves the spatial, temporal, and historical coordination information captured by the TGN while incorporating higher-level information regarding the latent behavioral role exhibited by the agent. The resulting role-enhanced representations are provided to the downstream task-specific decoders, whose learning objectives jointly optimize the MR-TGN encoder and the global role prototypes through end-to-end training.

\subsection{Sequential Multi-Task Decoder}
\begin{figure*}[t]
    \centering
    \includegraphics[width=0.9\textwidth]{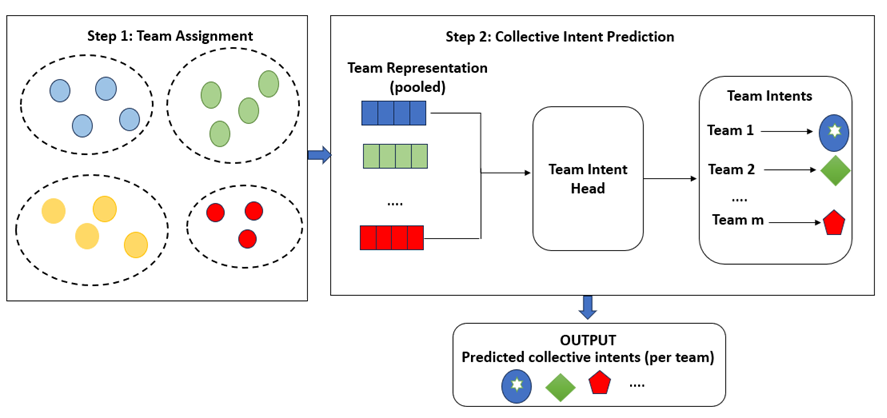}
    \caption{Overall architecture of the proposed sequential multi-task Decoder.}
    \label{fig:decoder}
\end{figure*}

The decoder is designed to progressively transform the role-enhanced agent representations produced by the MR-TGN encoder into team-level collective intent predictions. As illustrated in Fig.~\ref{fig:decoder}, the decoder consists of two sequential task-specific stages: \textit{team assignment} and \textit{collective intent prediction}. The first stage identifies the latent team structure among interacting agents, while the second stage exploits the inferred team composition to construct team-level representations and predict the collective intent of each team. Although the two tasks are performed sequentially in the forward computation, the entire framework is trained jointly in an end-to-end manner, allowing the downstream learning objectives to optimize the MR-TGN encoder, global role prototypes, and task-specific decoder components.

\subsubsection*{\textit{Step 1: Team Assignment}}
Given the behavioral agent representations $\mathbf{z}_{i}^{t}$ produced by the Temporal Graph Network, the team assignment module estimates the associations among agents based on their learned spatio-temporal interaction representations. The objective of this stage is to identify groups of agents exhibiting compatible role, behavioral, and coordination characteristics without assuming a fixed team composition. Accordingly, the team assignment module produces a predicted partition of the agents into $M_t$ teams as $\mathcal{C}^{t}
= \left\{
\mathcal{C}_{1}^{t},
\mathcal{C}_{2}^{t},
\ldots,
\mathcal{C}_{M_t}^{t}
\right\}$, 
where $\mathcal{C}_{m}^{t}$ denotes the set of agents assigned to the $m^{th}$ team at time $t$, and $M_t$ denotes the number of inferred teams. Since the team structure is derived from the learned agent representations, the framework can accommodate variations in team composition and interaction configurations across different scenarios.

\subsubsection*{Step 2: Collective Intent Prediction}
The inferred team assignments are subsequently used to construct team-level representations for collective intent prediction. For each inferred team $\mathcal{C}_{m}^{t}$, the role-enhanced representations of its constituent agents are aggregated using a permutation-invariant pooling operation:
\begin{equation}
\mathbf{u}_{m}^{t}
=
\operatorname{Pool}
\left(
\left\{
\mathbf{h}_{i}^{t}
\mid
i \in \mathcal{C}_{m}^{t}
\right\}
\right),
\label{eq:team_representation}
\end{equation}
where $\operatorname{Pool}(\cdot)$ denotes the team-level aggregation operation and $\mathbf{u}_{m}^{t}$ represents the resulting team embedding. The pooling operation enables the construction of a fixed-dimensional team representation independent of the number and ordering of agents within the team.

The resulting team representation is provided to the team intent prediction head, which estimates the probability distribution over the possible collective intent classes:
\begin{equation}
\hat{\mathbf{y}}_{m}^{t}
=
\operatorname{softmax}
\left(
f_{\mathrm{intent}}
\left(
\mathbf{u}_{m}^{t}
\right)
\right),
\label{eq:team_intent_prediction}
\end{equation}
where $f_{\mathrm{intent}}(\cdot)$ denotes the learnable team intent prediction head and $\hat{\mathbf{y}}_{m}^{t}$ denotes the predicted collective intent distribution of the $m^{th}$ team. The final output of the decoder is a set of team-level collective intent predictions given by $\hat{\mathcal{Y}}^{t}
=
\left\{
\hat{\mathbf{y}}_{1}^{t},
\hat{\mathbf{y}}_{2}^{t},
\ldots,
\hat{\mathbf{y}}_{M_t}^{t}
\right\}$, 
where each prediction corresponds to one of the teams identified by the team assignment stage. By explicitly performing team structure discovery before collective intent prediction, the proposed decoder enables team-aware reasoning over dynamic multi-agent systems and avoids directly inferring collective behaviors from an undifferentiated set of interacting agents.

\subsubsection*{Effect of Role Granularity on Early Situation Understanding}
The number of role prototypes determines the granularity of latent coordination patterns captured by the shared role memory. A small number of prototypes limits role diversity and causes different behaviors to be merged, whereas an excessive number of prototypes leads to fragmented representations and reduced generalization. Moderate values of $K$ provide an appropriate balance between representational capacity and robustness, resulting in improved collective intent discrimination, early correct recognition, and faster prediction stabilization. Experimental results indicate that moderate values of $K$ provide the best trade-off between representational capacity and robustness, leading to early and more stable intent recognition. A detailed sensitivity analysis of the effect of K is presented in Section \ref{sec:sensitivity}

\section{Case Study}
\label{sec:case_study}
The proposed framework is evaluated through a case study involving dynamic multi-agent interactions, with the objective of assessing its ability to discover evolving team structures and predict team-level collective intents. Air combat is considered for the case study due to its highly dynamic interaction patterns, heterogeneous agent behaviors, evolving team compositions, and coordinated mission execution. These characteristics provide a challenging multi-agent environment for evaluating the capability of the proposed framework to learn temporal coordination patterns, discover latent behavioral roles, identify team structures, and predict collective intents. Accordingly, diverse air combat scenarios are simulated to generate temporal interaction data containing agent-level state evolution and pairwise coordination information. The proposed framework is trained and evaluated on the resulting dataset to analyze its performance in team assignment and collective intent prediction. The remainder of this section first describes the dataset generation procedure,  followed by the experimental results and discussion.
\subsection{Dataset Generation}
Existing publicly available multi-agent datasets are primarily designed for tasks such as trajectory prediction, interaction modeling, or individual behavior analysis, and therefore do not provide the annotations and scenario diversity required for collective intent prediction in heterogeneous teams. In particular, they typically lack explicit team-level intent labels, dynamic team compositions, mission-specific coordination patterns, and latent role transitions that are essential for evaluating the proposed framework. Consequently, a dedicated dataset was generated to provide complete control over mission design, agent behaviors, role assignments, and intent annotations while ensuring the availability of synchronized temporal interaction data required for training and evaluation. The generated dataset has been made publicly available through the \href{https://github.com/Nagaranikavin/Heterogeneous-Team-Intent-Dataset}{Heterogeneous-Team-Intent-Dataset} GitHub repository.

The air combat scenarios are generated using SCRIMMAGE, which stands for Simulating Collaborative Robots in Massive Multi-Agent Game Execution, a modular simulation framework designed for the development and evaluation of autonomous multi-agent systems \cite{kevin2019SCRIMMAGE}. SCRIMMAGE supports the simulation of multiple interacting agents with configurable motion models, behaviors, and mission-level logic within a common simulation environment. Its modular architecture enables the construction of diverse multi-agent scenarios and the systematic generation of time-varying agent trajectories and state information. In this study, SCRIMMAGE is employed to construct air combat scenarios with varying mission compositions, agent roles, and behavioral evolution. Mission-specific behaviors and phase transitions \cite{Zhao2025UAVSwarmRL} are implemented to generate heterogeneous and homogeneous multi-agent interaction patterns corresponding to the considered collective intents. During each simulation run, the time-varying states of all aircraft are recorded and subsequently processed to construct temporal interaction graphs for training and evaluating the proposed framework. The use of a common simulation environment enables controlled generation of diverse scenarios while systematically varying team compositions and mission configurations.

The simulation environment comprises multiple aircraft exhibiting either heterogeneous or homogeneous motion characteristics, temporally evolving behaviors, and coordinated mission execution. This case study is designed to evaluate the proposed framework's ability to identify dynamically formed teams and infer their collective intents from spatio-temporal interaction patterns. The simulation considers five collective intent labels: Sky Sentinel Strike (SSS), Escort Penetration Strike (EPS), Anti-Ship Strike (ASS), Reconnaissance (Recon), and No Mission (NM). The first four intent labels correspond to coordinated mission behaviors, whereas the NM label represents aircraft that do not exhibit the characteristic behavioral and coordination patterns associated with any of the modeled missions. Importantly, individual simulation scenarios do not necessarily contain all mission types simultaneously. Instead, the generated scenarios consist of different combinations of aircraft assigned to one or more missions and aircraft not engaged in any mission, producing variations in the number of teams, team compositions, agent behaviors, and interaction configurations.

The four coordinated missions are designed to capture different forms of collective behavior. SSS, EPS, and ASS consist of heterogeneous agents exhibiting distinct motion characteristics according to their functional roles within the team. In contrast, the Recon mission consists of homogeneous agents that initially exhibit similar behavioral characteristics and collectively transition through different phases of mission execution. Inspired by recent studies that model UAV swarm operations as multi-stage collaborative tasks with distinct objectives across mission phases, our missions are also organized into sequential operational phases that represent the evolution of team behavior during mission execution \cite{Zhao2025UAVSwarmRL}. This combination of heterogeneous and homogeneous teams provides a challenging evaluation setting, since team identification and intent prediction cannot rely solely on similarities in individual agent trajectories. The five collective intents are briefly described below.

\subsubsection*{Sky Sentinel Strike (SSS)}
The SSS mission consists of four agents with distinct behavioral roles: a striker, a left flanker, a right flanker, and a decoy. The mission is inspired by coordinated unmanned combat aerial vehicles (UCAV) air combat strategies, where multiple aircraft cooperate through role-dependent tactical coordination to accomplish a common engagement objective \cite{Liu2025TacticalCoordination}.During the initial phase, the agents move toward the operational region while maintaining coordination as a team. In the second phase, the agents begin to exhibit role-dependent motion characteristics. The striker proceeds along the primary approach direction, while the left and right flankers deviate toward different sides of the operational region. The decoy follows a distinct maneuver pattern intended to create behavioral separation from the primary attacking agent. During the final phase, the agents execute their respective engagement behaviors. The striker proceeds toward the target and, upon reaching the Weapon Engagement Zone (WEZ), initiates weapon engagement. Meanwhile, the left and right flankers maintain their respective approach directions and engage the target from spatially separated positions, while the decoy continues its distinct maneuver pattern. This phase therefore exhibits simultaneous role-dependent engagement behaviors associated with the same collective mission. Consequently, the SSS mission produces a heterogeneous team in which agents belonging to the same collective mission exhibit substantially different individual motion patterns over time.

\subsubsection*{Escort Penetration Strike (EPS)}
The EPS mission consists of one primary mission agent and three escort agents. The mission is motivated by UAV penetration operations, in which an aircraft progresses through adversarial environments to reach and engage a protected target \cite{Xing2024UAVPenetration}. Unlike the single-UAV penetration scenario considered in \cite{Xing2024UAVPenetration}, the proposed mission models a coordinated team in which a primary penetration agent is supported by multiple escort agents that maintain formation and provide coordinated protection throughout the mission execution. During the first phase, the agents approach the operational region while exhibiting relatively coherent team motion. In the second phase, the primary agent continues progressing toward the mission objective, whereas the escort agents adjust their trajectories according to their supporting roles and the evolving team configuration. During the final phase, the primary mission agent proceeds toward the target and initiates weapon engagement upon reaching the WEZ. Meanwhile, the escort agents maintain a certain distance from the target and perform coordinated engagement behaviors while preserving their supporting configuration around the primary mission agent. Thus, the agents exhibit distinct role-dependent behaviors while collectively executing the same mission intent. The EPS mission therefore represents a heterogeneous team whose members exhibit different motion characteristics while remaining associated with the same collective intent.

\subsubsection*{Anti-Ship Strike (ASS)}
The ASS mission comprises four heterogeneous agents: a low-altitude attacker, a left flanker, a right flanker, and a decoy. The mission is motivated by anti-ship strike operations, where a primary attacker follows a carefully planned low-altitude, multi-stage approach trajectory to penetrate shipborne defenses before engaging the target \cite{Sun2025HybridMOPSO}. During the initial phase, the agents approach the operational region as a coordinated group. In the second phase, the agents begin to exhibit role-specific behavioral patterns. The low-altitude attacker transitions toward a lower-altitude approach trajectory, while the flankers maneuver along different lateral directions and the decoy exhibits a distinct trajectory pattern. During the final phase, the low-altitude attacker continues its approach toward the target and initiates weapon engagement upon reaching the WEZ. Meanwhile, the left and right flankers maintain spatially separated positions and perform engagement behaviors toward the target, while the decoy continues its distinct maneuver pattern. This phase therefore exhibits heterogeneous role-dependent engagement behaviors associated with the same collective mission. The simultaneous presence of altitude-dependent, lateral, and diversionary motion characteristics introduces substantial intra-team behavioral diversity and provides a challenging setting for discovering team membership from temporal interaction patterns.

\subsubsection*{Reconnaissance (Recon)}
The Recon mission consists of homogeneous agents that share similar behavioral capabilities and collectively execute a sequence of temporally evolving activities. During the first phase, the agents explore the operational region to search for the target. Their trajectories may spatially diverge during this phase to increase the explored area. Once the target is detected, the mission transitions to the second phase, during which the agents converge toward the detected target region. In the final phase, the agents perform coordinated target-tracking behavior by maintaining observation of the target while adapting their trajectories according to its movement. This phased behavior is motivated by recent studies on multistage UAV swarm missions involving distributed target search and subsequent information aggregation \cite{Zhao2025UAVSwarmRL}. Unlike the other missions, the Recon agents are not differentiated through predefined heterogeneous functional roles. Instead, the collective intent is characterized primarily by the shared temporal evolution from exploration to convergence and subsequent tracking. This mission therefore evaluates whether the proposed framework can recognize collective intent from temporally evolving coordination patterns among homogeneous agents.

\subsubsection*{No Mission (NM)}
In addition to the four coordinated missions, the case study also includes aircraft that are not engaged in any of the modeled missions. These aircraft constitute the NM class and follow trajectories that do not exhibit the structured role differentiation or phase-dependent coordination patterns associated with SSS, EPS, ASS, or Recon. The inclusion of the NM class prevents the prediction task from being restricted to distinguishing only among predefined coordinated missions and requires the model to differentiate mission-specific collective behavior from agents exhibiting non-mission motion patterns.

The coordinated missions evolve through three temporal phases representing the progression from initial movement, through mission-specific behavioral evolution, to the final mission execution stage. Although the semantic interpretation of these phases differs across missions, the phased scenario design enables the generation of temporally evolving interaction patterns and gradual behavioral transitions. This is particularly important for evaluating the proposed temporal framework, since the collective intent of a team may not be immediately distinguishable from early observations and becomes increasingly evident as agent behaviors and coordination patterns evolve over time.

All scenarios are simulated using SCRIMMAGE, from which the time-varying states and trajectories of the aircraft are recorded. Different simulation scenarios contain different combinations of coordinated missions and NM aircraft rather than including all five intent labels in every scenario. Consequently, the generated dataset exhibits variations in the number of active teams, team composition, heterogeneous and homogeneous agent behaviors, and interactions among agents belonging to different collective intents. These variations provide a challenging evaluation environment for assessing the generalization capability of the proposed framework under diverse multi-agent interaction configurations.
In the considered multi-agent environment, the node feature vector is defined as
\begin{equation}
\mathbf{x}_{i}^{t}
=
\left[\
\mathbf{p}_{i}^{t},
\mathbf{v}_{i}^{t},
s_i^{t},
\theta_i^{t},
\omega_i^{t},
a_i^{t}
\right]
\label{eq:node_features}
\end{equation}
where $\mathbf{p}_{i}^{t} = [x_i^{t}, y_i^{t}, z_i^{t}]$ denotes the three-dimensional position of agent $v_i$, and $\mathbf{v}_{i}^{t} = [v_{x,i}^{t}, v_{y,i}^{t}, v_{z,i}^{t}]$ represents its three-dimensional velocity. The scalar $s_i^{t}$ denotes the speed of the agent, $\theta_i^{t}$ represents its heading angle, $\omega_i^{t}$ denotes the turn rate characterizing the temporal variation of the heading, and $a_i^{t}$ represents the acceleration of the agent.

For every interacting pair of agents $(v_i,v_j)$, an edge feature vector $\mathbf{e}_{ij}^{t}
\in \mathbb{R}^{d_e}$ is constructed to characterize their relative spatial and motion relationships. The edge feature vector is defined as
\begin{equation}
\mathbf{e}_{ij}^{t}
=
\left[
d_{ij}^{t},
\Delta \mathbf{p}_{ij}^{t},
\Delta \mathbf{v}_{ij}^{t},
\Delta s_{ij}^{t},
\Delta z_{ij}^{t},
\Delta \theta_{ij}^{t},
v_{c,ij}^{t}
\right].
\end{equation}
where $d_{ij}^{t}$ denotes the relative distance between agents $v_i$ and $v_j$, $\Delta \mathbf{p}_{ij}^{t}$ represents their relative position vector, and $\Delta \mathbf{v}_{ij}^{t}$ denotes their relative velocity vector. The scalar $\Delta s_{ij}^{t}$ represents the difference in speed between the two agents, while $\Delta z_{ij}^{t}$ denotes their relative altitude. Furthermore, $\Delta \theta_{ij}^{t}$ represents the difference in heading, and $v_{c,ij}^{t}$ denotes the closing velocity, which characterizes the rate at which the distance between the two agents changes over time and indicates whether the agents are converging toward or diverging away from each other. 

For the air combat case study, the behavioral node features are selected directly from the node feature vector defined in \ref{eq:node_features}. Specifically, the agent's three-dimensional position, velocity, speed, heading angle, turn rate, and acceleration are used to characterize its intrinsic motion dynamics and behavioral evolution over time. These features describe the temporal behavior of each agent independently and exclude explicit relational information.

\subsection{Results and Discussion}
This section presents the experimental results and evaluates the proposed framework in terms of overall predictive performance, convergence behavior, early prediction capability, and temporal prediction stability. The performance of the proposed MR-TGN is first compared with the baseline models using conventional learning metrics, including validation accuracy and training loss. 
Subsequently, the temporal characteristics of intent recognition are investigated using the Early Correct Sample (ECS) and Time To Stable Sample at the $k^{th}$ timestep (TTSS@k) metrics. This evaluation strategy enables the proposed framework to be assessed not only in terms of its final predictive performance, but also in terms of how quickly and consistently correct intent predictions can be obtained from partially observed interaction sequences.

\subsubsection{Experimental setup}
The proposed MR-TGN model was implemented using the PyTorch framework. Table \ref{tab:hyperparameters} presents the implementation details and hyperparameter settings used to obtain the reported experimental results. Unless otherwise specified, the same configuration was maintained throughout all experiments to ensure a consistent and reproducible evaluation. The influence of the latent prototype parameter ($K$) is analyzed separately in the \ref{sec:sensitivity} section.

\begin{table}[!t]
\caption{Implementation details and hyperparameter settings.}
\label{tab:hyperparameters}
\centering
\begin{tabular}{lc}
\hline
\textbf{Parameter} & \textbf{Value} \\
\hline
Optimizer & Adam \\
Learning rate & 0.001 \\
Temporal window length & 20 timestamps \\
Number of epochs & 100 \\
Node embedding dimension & 32 \\
Memory embedding dimension & 32 \\
Number of attention heads & 4 \\
Dropout rate & 0.3 \\
Number of latent prototypes ($K$) & 8 \\
\hline
\end{tabular}
\end{table}

The proposed framework is compared with several representative spatio-temporal learning models, including Graph Attention Networks (GAT), Attention-LSTM, GNN+Bi-LSTM, Spatio-Temporal Graph Convolutional Networks (STGCN), Dynamic Self-Attention Networks (DySAT), a TGN-based baseline, and the proposed MR-TGN. Wherever applicable, the baseline encoders are integrated into a common prediction framework to ensure a fair comparison by isolating the contribution of the representation learning module from that of the prediction heads. For interaction-aware models, the same team assignment and intent prediction heads are employed. In contrast, models that learn only individual temporal representations, such as Attention-LSTM, do not explicitly capture inter-agent relationships and therefore are evaluated only for the intent prediction task.

These baseline models were selected because they represent the major categories of sequence and graph-based learning approaches for spatio-temporal modeling. GAT and Attention-LSTM each capture only one aspect of the problem. GAT effectively models spatial relationships through graph attention mechanisms but does not explicitly capture long-term temporal evolution [\cite{Zhang2020A, Sun2023AttentionSurvey}. Conversely, Attention-LSTM learns temporal dependencies using attention-enhanced recurrent networks but lacks an explicit mechanism to model inter-agent spatial interactions[\cite{Mavsar2024GANLSTM}. GNN+Bi-LSTM and STGCN both integrate spatial and temporal information for spatio-temporal representation learning. However, they generally rely on static or predefined graph structures, limiting their ability to model continuously evolving interactions among agents in highly dynamic multi-agent environments. While GNN+Bi-LSTM learns spatial and temporal dependencies in separate stages dynamics [\cite{Vitulyova2025A}], STGCN performs joint spatio-temporal graph convolutions [\cite{Luo2024STMGCN,Siabi2026Innovative}], yet both assume relatively stable graph connectivity. DySAT employs temporal self-attention to learn evolving graph representations \cite{Sankar2020DySAT:}]  but primarily operates on graph snapshots without maintaining persistent node memory. The TGN baseline  addresses this limitation through a memory module that continuously updates node representations over time, enabling effective continuous-time temporal learning [\cite{rossi2020temporalgraphnetworksdeep, Jeon2025Leveraging}]. However, it learns generic node representations without explicitly capturing functional role similarities among heterogeneous agents. In contrast, the proposed MR-TGN extends TGN with meta-role representation learning, enabling agents with similar behavioral functions to share transferable knowledge while preserving temporal interaction dynamics, thereby improving both team assignment and collective intent prediction.

All reported results are evaluated on the held-out test set comprising 
multiple independently generated scenarios. Unless otherwise stated, the 
reported team assignment (TA) and intent prediction (IP) accuracies are dataset-level metrics obtained by 
aggregating predictions from all eligible temporal graph states across 
the test scenarios, rather than results obtained from any individual 
scenario. For the phase-wise evaluation, predictions belonging to the 
same mission phase are pooled across all test scenarios, and the 
corresponding accuracy is computed over the resulting set of predictions. Similarly, the results under the full-history and partial-observation settings are computed using all eligible predictions from the complete test set under the respective observation protocol. The same fixed scenario-level train, validation, and test partitions are used for all compared methods to ensure a consistent and fair~evaluation.

\subsubsection{Predictive Performance Evaluation}
\begin{figure*}[t]
    \centering
    \includegraphics[width=0.8\textwidth]{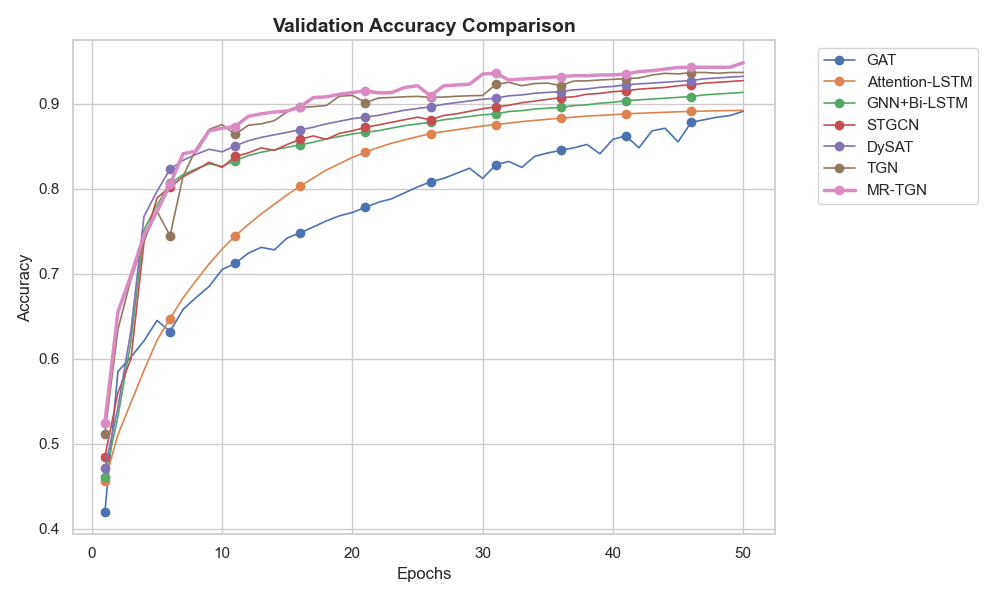}
    \caption{Validation Accuracy for the proposed MR-TGN and baseline models.}
    \label{fig:validation accuracy}
\end{figure*}

Fig. \ref{fig:validation accuracy} presents the variation of validation accuracy of the evaluated models over 50 training epochs. The results reveal a progressive improvement in predictive performance as the modeling capability advances from isolated spatial or temporal learning toward joint spatio-temporal interaction modeling, persistent historical memory, and latent role-aware reasoning. GAT and Attention-LSTM exhibit comparatively lower validation performance because they model only one aspect of the multi-agent dynamics. GNN+Bi-LSTM improves performance by combining spatial encoding with sequential temporal learning, while STGCN provides stronger modeling of local spatio-temporal coordination patterns. However, GNN+LSTM limits direct modeling of evolving coordination due to its factorized spatial and temporal processing, while STGNN is constrained by its bounded temporal receptive field and lack of persistent memory for capturing long-range historical dependencies. DySAT achieves higher validation accuracy by combining structural attention over graph snapshots with temporal self-attention across evolving node representations. TGN further improves performance through persistent node-level memory, which preserves historical interaction states and allows current agent representations to depend on previously observed interactions and maneuver evolution. This capability is particularly important in multi-phase multi-agent scenarios, where similar instantaneous configurations may correspond to different intents depending on the preceding interaction history. In addition to this, the proposed meta-role memory learns globally shared, task-optimized latent behavioral prototypes that capture recurring within-agent behavioral dynamics associated with functional roles. 

\begin{table*}[!t]
\centering
\caption{Phase-wise TA and IP accuracy over all eligible predictions in the held-out test scenarios under the full-history observation setting.}
\label{tab:phase_full_history}
\small
\begin{tabular}{lcccccccc}
\hline
\multirow{2}{*}{Model}
& \multicolumn{2}{c}{Phase 1}
& \multicolumn{2}{c}{Phase 2}
& \multicolumn{2}{c}{Phase 3}
& \multicolumn{2}{c}{Overall} \\
\cline{2-9}
& TA Acc. & IP Acc.
& TA Acc. & IP Acc.
& TA Acc. & IP Acc.
& TA Acc. & IP Acc. \\
\hline

GAT
& 95.0 & 70.1
& 89.2 & 81.5
& 82.7 & 85.6
& 89.0 & 79.1 \\

Attention-LSTM
& -- & 69.4
& -- & 85.3
& -- & 90.8
& -- & 81.8 \\

GNN+LSTM 
& 93.1 & 70.8
& 90.8 & 86.9
& 89.5 & 91.7
& 91.1 & 83.1 \\

STGNN
& 94.8 & 73.6 & 92.5 & 88.4 & 91.8 & 94.1 & 93.0 & 84.8 \\

DySAT
& 91.7 & 74.2
& 93.0 & 90.6
& 93.5 & 95.2
& 92.8 & 89.3 \\

TGN
& 95.2 & 75.6
& 95.1 & 92.8
& 95.5 & 96.2
& 95.3 & 91.5 \\

\textbf{MR-TGN}
& \textbf{96.3} & \textbf{78.9}
& \textbf{96.8} & \textbf{94.7}
& \textbf{97.1} & \textbf{97.4}
& \textbf{96.7} & \textbf{92.7} \\
\hline
\end{tabular}%

\footnotesize{\textit{Note:} "--" indicates that the metric is not applicable, as Attention-LSTM models agents independently and does not support team assignment.}
\end{table*}

\begin{table*}[!t]
\centering
\caption{Phase-wise TA and IP accuracy over all eligible predictions in the held-out test scenarios when observation begins from Phase 2.}
\label{tab:phase_partial_history}
\small
\begin{tabular}{lcccccc}
\hline
\multirow{2}{*}{Model}
& \multicolumn{2}{c}{Phase 2}
& \multicolumn{2}{c}{Phase 3}
& \multicolumn{2}{c}{Overall} \\
\cline{2-7}
& TA Acc. & IP Acc.
& TA Acc. & IP Acc.
& TA Acc. & IP Acc. \\
\hline

GAT
& 84.1 & 76.2
& 79.5 & 82.8
& 81.8 & 79.5 \\

Attention-LSTM
& -- & 77.8
& -- & 83.9
& -- & 80.8 \\

GNN+LSTM
& 86.2 & 81.3
& 86.9 & 88.1
& 86.6 & 84.7 \\

STGNN
& 87.4 & 82.6
& 87.8 & 88.9
& 87.6 & 85.8 \\

DySAT
& 88.1 & 84.4
& 90.2 & 92.3
& 89.2 & 88.4 \\

TGN
& 90.8 & 87.5
& 93.1 & 94.0
& 92.0 & 90.8 \\

\textbf{MR-TGN}
& \textbf{93.5} & \textbf{90.6}
& \textbf{95.4} & \textbf{93.1}
& \textbf{94.5} & \textbf{91.4} \\

\hline
\end{tabular}%

\footnotesize{\textit{Note:} "--" indicates that the metric is not applicable, as Attention-LSTM models agents independently and does not support team assignment.}
\end{table*}

Tables~\ref{tab:phase_full_history} and~\ref{tab:phase_partial_history} analyze the team assignment and intent prediction performance at different stages of mission progression. Under the full-history observation setting, a clear difference can be observed between the difficulty of team assignment and intent prediction during the early mission phase. In Phase~1, most agents move toward the operational region and maintain relatively consistent spatial and kinematic relationships with their teammates. Features such as relative distance, heading alignment, and velocity correlation therefore provide sufficient relational evidence for identifying agents that move as part of the same team. Consequently, all models achieve comparatively high team-assignment accuracy during this phase. In contrast, intent prediction is considerably more difficult in Phase~1. The early ingress behaviors of different missions may exhibit substantial similarities, since the agents have not yet initiated their mission-specific tactical maneuvers. For example, agents belonging to strike, escort, and anti-ship missions may all exhibit coordinated movement toward the operational region. Although these observations provide evidence regarding team membership, they contain limited information about the subsequent tactical objective of the team. This explains the substantial difference between team-assignment and intent-prediction accuracy during Phase~1. 

During Phase~2, agents begin to separate according to their tactical roles, producing mission-dependent changes in their spatial and maneuver relationships. The evolution of relative distance, heading alignment, velocity correlation, together with relative turn-rate and acceleration differences, provides increasingly discriminative evidence of coordinated role separation. Consequently, intent-prediction accuracy improves substantially across the models. However, instantaneous spatial models remain limited because they infer the mission primarily from the current relational configuration. Temporal models can additionally exploit the preceding ingress behavior and the evolution of interactions leading to role separation, resulting in improved intent~recognition.

Phase~3 corresponds to mission-specific execution, such as attack, tracking, escort support attack, or coordinated maritime strike. Although the current agent behaviors become more distinctive, instantaneous observations may still be ambiguous because multiple missions contain attack-related maneuvers and heterogeneous agents belonging to the same team may exhibit strongly divergent trajectories. Temporal models achieve higher performance because sufficient interaction history has accumulated by this stage. In particular, TGN retains the historical evolution of pairwise interactions from the earlier coordinated ingress and role-separation phases, enabling the current role-dependent behaviors to be interpreted in their preceding relational context.
MR-TGN consistently achieves the highest team-assignment and intent-prediction accuracy across the three phases. The improvement over TGN indicates that temporal interaction memory alone does not fully capture the recurring behavioral characteristics associated with heterogeneous tactical roles. The meta-role memory provides additional role-aware behavioral context, enabling the model to distinguish agents exhibiting different tactical behaviors while preserving their association with the same coordinated team. This complementary modeling of historical relational evolution and latent role-dependent behavior results in reliable team assignment and intent prediction throughout mission~progression. Attention-LSTM is evaluated only for the intent prediction task because it models each agent independently through temporal sequence learning without explicitly capturing inter-agent interactions or team-level representations. Consequently, it does not produce the relational embeddings required for the team assignment task, and the corresponding entries are marked as not applicable in Table~\ref{tab:phase_full_history}, \ref{tab:phase_partial_history}.

Table~\ref{tab:phase_partial_history} presents the phase-wise team assignment and intent prediction performance under the partial-history observation setting, where agents become observable only from Phase~2 and no information from the initial ingress phase is available. This setting is considerably more challenging because the first available observations correspond to an ongoing role-separation stage, where agents belonging to the same team already exhibit heterogeneous and increasingly divergent behaviors. Consequently, the models must infer team membership and mission intent without observing the preceding coordinated evolution that established the underlying interaction patterns.

GAT relies solely on the current graph snapshot and therefore infers team relationships from instantaneous node and edge features. When observation begins in Phase~2, coordinated agents may already differ significantly in heading, velocity, altitude, and maneuver characteristics due to role specialization. Without historical interaction information, these instantaneous differences can be difficult to distinguish from unrelated behaviors, leading to reduced team assignment and intent prediction accuracy. Attention-LSTM also experiences reduced intent prediction accuracy since the shortened observation window provides less temporal evidence for learning mission-specific behavioral evolution. Because it models agents independently, no team assignment results are reported.

The performance degradation of GNN+LSTM and \linebreak STGNN is comparatively smaller, as both models continue to exploit the available spatio-temporal observations after Phase~2. However, the missing ingress phase removes part of the temporal context that characterizes the transition from coordinated formation to tactical role separation, thereby limiting their ability to fully capture mission progression. DySAT demonstrates greater robustness under partial observation by selectively attending to the most informative spatial and temporal interactions within the observed interval. Nevertheless, the absence of the initial coordination phase still reduces the available temporal evidence, leading to a moderate decline compared with the full-history setting. Compared with the full-history setting, TGN exhibits a moderate performance degradation bec MR-TGN shows the smallest performance degradation among all evaluated models. Even when observations begin from Phase~2, the learned latent behavioral prototypes complement the accumulated temporal interactions by associating partially observed behaviors with higher-level role representations. This enables the model to recover team associations and mission intent more effectively despite the missing ingress phase, demonstrating its robustness under incomplete observation.ause the initial team interaction context established during Phase~1 is unavailable. Nevertheless, it consistently outperforms DySAT, indicating that the accumulated interaction history from the observed phases remains effective for team assignment and intent prediction. MR-TGN shows the smallest performance degradation among all evaluated models. Even when observations begin from Phase~2, the learned latent behavioral prototypes complement the accumulated temporal interactions by associating partially observed behaviors with higher-level role representations. This enables the model to recover team associations and mission intent more effectively despite the missing ingress phase, demonstrating its robustness under incomplete observation.

\subsubsection{Temporal Performance Evaluation Using Early and Stability Metrics}
While conventional classification metrics provide an overall assessment of model performance, they do not fully characterize the temporal behaviour of predictions in dynamic multi-agent environments. In such environments, the practical effectiveness of a prediction framework depends not only on whether the underlying intent is correctly identified, but also on how early a reliable prediction can be obtained and whether the predicted intent remains stable over subsequent observations. Therefore, in addition to conventional performance measures, ECS and TTSS@k are adopted to assess the temporal characteristics of the proposed framework and provide a more comprehensive evaluation of its prediction capability.

Early and stable prediction is particularly important in safety-critical and time-sensitive multi-agent applications such as air combat, autonomous surveillance, cooperative robotics, and intelligent transportation systems. In an air combat scenario, for example, an agent's intent may evolve progressively through its interactions, motion patterns, and coordination with other agents. Waiting until a complete trajectory is observed before identifying the intent may significantly reduce the time available for threat assessment, tactical planning, and appropriate countermeasures. Consequently, a model capable of identifying the correct intent from partial observations is more operationally useful than a model that achieves high classification accuracy only after observing most of the trajectory. Moreover, an isolated correct prediction at an early sample does not necessarily indicate reliable intent recognition, since predictions may fluctuate as additional observations become available. Hence, both early prediction capability and temporal stability are considered in the evaluation.

\begin{figure*}[t]
    \centering
    \includegraphics[width=0.9\textwidth]{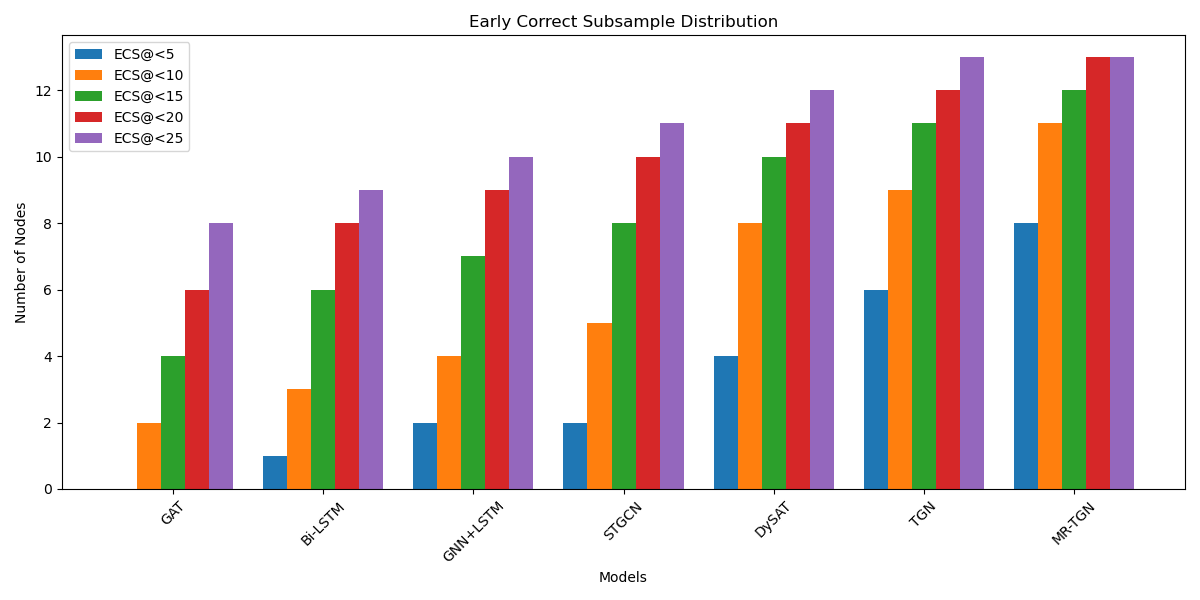}
    \caption{Cumulative number of agents achieving correct intent prediction within different ECS thresholds for the proposed MR-TGN and baseline models.}
    \label{fig:ECS}
\end{figure*}

The ECS metric measures the earliest sample at which the predicted intent matches the ground-truth intent. For a trajectory containing a sequence of predictions, ECS identifies the first occurrence of a correct prediction, irrespective of whether the subsequent predictions remain correct. A lower ECS value therefore indicates that the model is capable of recognizing the underlying intent using fewer temporal observations. This metric provides a direct measure of the early prediction capability of the model and enables the comparison of different methods in terms of how quickly meaningful intent information can be extracted from evolving agent interactions. Fig. \ref{fig:ECS} presents the cumulative distribution of ECS values obtained by the proposed MR-TGN and the baseline models at different observation thresholds. The results demonstrate a progressive improvement in early intent recognition from conventional spatial and sequential models to spatio-temporal graph-based approaches. Among the compared methods, MR-TGN achieves the highest number of correctly predicted nodes at the earliest observation thresholds. Specifically, MR-TGN correctly identifies the intent of 8 nodes within the first five samples and 11 nodes within the first ten samples, compared with 6 and 9 nodes, respectively, for TGN. In contrast, several baseline models, particularly GAT, fail to correctly recognize any node within the first five samples and require considerably longer observation sequences to obtain correct predictions. Furthermore, MR-TGN correctly predicts the intent of 12 nodes within 15 samples and all 13 nodes within 20 samples, whereas the remaining methods require longer observation horizons or fail to achieve comparable early recognition performance. These results indicate that the proposed meta-role-based temporal representation enables the model to extract discriminative intent information from shorter interaction histories, thereby improving its suitability for time-sensitive multi-agent decision-making scenarios.
\begin{figure*}[t]
    \centering
    \includegraphics[width=0.9\textwidth]{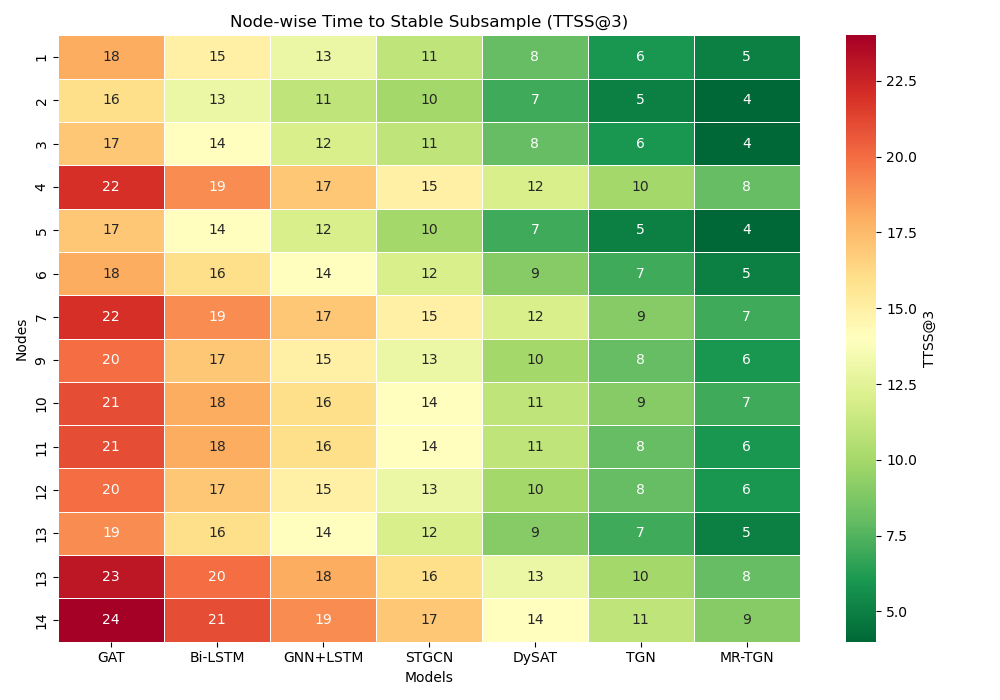}
    \caption{Node-wise comparison of TTSS@3 values for the proposed MR-TGN and baseline models}
    \label{fig:TTSS}
\end{figure*}
However, since ECS considers only the first correct prediction, it does not account for subsequent prediction fluctuations and therefore cannot independently characterize the temporal reliability of the predicted intent. 

To complement ECS, TTSS@k is employed to evaluate how quickly the model produces consistently correct predictions. TTSS@k identifies the earliest sample at which the model generates $k$ consecutive correct predictions. In this work, $k=3$ is adopted, and the corresponding metric is denoted as TTSS@3. Specifically, TTSS@3 determines the earliest sample from which three consecutive predictions agree with the ground-truth intent. Compared with ECS, this metric imposes a stronger temporal consistency requirement and reduces the influence of isolated or transient correct predictions. 
Fig.~\ref{fig:TTSS} presents the node-wise TTSS@3 values obtained by the proposed MR-TGN and the baseline methods. The heatmap reveals clear differences in the time required by different models to establish temporally consistent intent predictions. Conventional graph-based and sequential models generally exhibit higher TTSS@3 values, indicating that a longer interaction history is required before stable predictions can be obtained. Although spatio-temporal graph models improve prediction stability by jointly exploiting relational and temporal information, the proposed MR-TGN consistently achieves lower TTSS@3 values across the evaluated nodes. This demonstrates that MR-TGN not only produces correct predictions at earlier observation stages, as indicated by the ECS results, but also converts these early predictions into temporally consistent predictions within a shorter observation horizon. The improvement can be attributed to the complementary contributions of temporal interaction modeling and meta-role representations, which enable the framework to preserve evolving interaction patterns while exploiting latent behavioral similarities among agents. Consequently, the proposed model provides early and stable intent prediction, an important characteristic for multi-agent applications in which downstream tactical decisions must be made before complete trajectories become available. 

The joint analysis of ECS and TTSS@3 provides complementary insights into the temporal prediction behaviour of the proposed framework.
A small difference between ECS and TTSS@3 indicates that the model stabilizes shortly after its first correct prediction, while a large difference suggests that the early predictions exhibit temporal fluctuations before consistent correct predictions are achieved. Therefore, evaluating both metrics enables a more comprehensive assessment of the suitability of the proposed framework for dynamic multi-agent environments.

\subsubsection{Sensitivity Analysis}
\label{sec:sensitivity}
The sensitivity of MR-TGN to the number of latent role prototypes was 
investigated by varying the role-memory capacity as 
$K \in \{2,4,6,8,10,12\}$. The influence of $K$ was evaluated using the ECS and $\mathrm{TTSS}@3$ metrics. Since lower values are preferred for both metrics, configurations whose observations are concentrated toward the lower-left region of the ECS--$\mathrm{TTSS}@3$ space indicate better temporal prediction performance. To ensure that the selection of the role-memory capacity was not influenced by scenario-specific trajectory variations, each value of $K$ was evaluated across multiple scenarios, and the final configuration was selected based on the aggregated ECS and $\mathrm{TTSS}@3$ performance over all evaluated scenarios.
\begin{figure*}[t]
    \centering
    \includegraphics[width=0.5\linewidth]{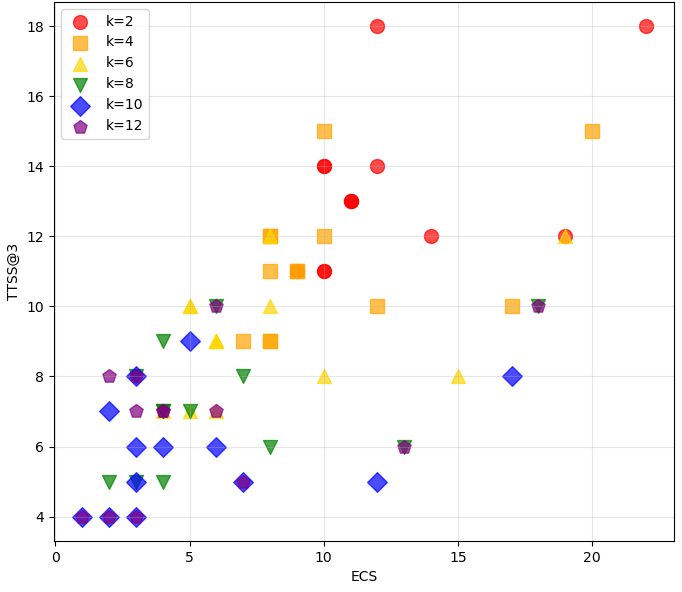}
    \caption{Sensitivity Analysis on the Number of Latent Role Prototypes}
     \label{fig:role_sensitivity}
\end{figure*}

As shown in Fig.~\ref{fig:role_sensitivity}, configurations with a small number of latent role prototypes, particularly $K=2$ and $K=4$, exhibit a wider dispersion of observations and several cases with high ECS and 
$\mathrm{TTSS}@3$ values. This indicates that a limited role-memory 
capacity is insufficient to represent the diversity of recurring 
behavioral patterns exhibited by heterogeneous agents, resulting in 
delayed correct recognition and reduced prediction stability. Increasing the number of role prototypes improves the temporal prediction performance, with the observations progressively shifting toward the 
lower-ECS and lower-$\mathrm{TTSS}@3$ region. Among the evaluated 
configurations, $K=8$ provides a favorable balance between early 
recognition and prediction stability, as most observations are concentrated 
at comparatively low values of both metrics. Increasing the role-memory 
capacity further to $K=10$ or $K=12$ does not provide a consistent 
improvement in either metric, indicating diminishing 
benefits from introducing additional latent role~prototypes.

These results suggest that an insufficient number of role prototypes 
limits the behavioral representation capacity of the global role memory, 
whereas excessively increasing the number of prototypes provides limited 
additional benefit and may introduce redundant latent behavioral 
representations. Therefore, $K=8$ was selected as the number of latent 
role prototypes for the remaining experiments, as it achieves a suitable 
trade-off between early intent recognition, temporal prediction stability, 
and model complexity.

\subsubsection{Ablation Study}
Table~\ref{tab:ablation} presents the ablation results obtained by 
systematically removing the major components of MR-TGN while retaining 
the same experimental settings. The full MR-TGN achieves the best 
performance, with an intent prediction accuracy of $92.7\%$, demonstrating that the proposed components provide complementary contributions to the overall framework. Removing the global role memory reduces the intent prediction accuracy to $90.72\%$. This degradation indicates that the globally shared role prototypes provide useful behavioral abstractions beyond the agent-specific temporal representations produced by the TGN encoder. Without node memory, the model cannot persistently accumulate and update information from previous interactions and agent states. Consequently, its predictions rely primarily on the currently available graph context and temporally local observations, limiting its ability to represent long-term behavioral evolution and historical dependencies. 
When the attention mechanism is removed, the model loses the ability to adaptively emphasize the neighboring interactions that are most informative for updating agent representations. This is particularly important in heterogeneous multi-agent environments, where different agents and pairwise interactions do not contribute equally to the evolving collective behavior. Finally, removing the team assignment head reduces the intent prediction accuracy to $89.43\%$. The resulting degradation indicates that the team-assignment objective acts as an intermediate structural learning task that encourages the encoder and meta-role representations to preserve information relevant to agent grouping and coordinated behaviors.

Overall, the ablation results demonstrate that the performance improvement of MR-TGN cannot be attributed to a single architectural component. Node memory provides persistent historical information, the attention mechanism selectively models interaction relevance, the global role memory captures transferable latent behavioral patterns, and the team assignment head encourages explicit learning of the underlying team structure. Among these components, node memory produces the largest performance reduction when removed, highlighting the importance of long-term historical dependencies for dynamic team-intent recognition. Nevertheless, the consistent degradation observed for all ablated variants confirms that the proposed components provide distinct and complementary contributions to the complete MR-TGN framework.

\begin{table}[htbp]
\centering
\caption{Ablation study results of the proposed MR-TGN model.}
\label{tab:ablation}
\begin{tabular}{lc}
\hline
\textbf{Variant} & \textbf{Intent Accuracy (\%)} \\
\hline
MR-TGN (Full)              & $92.73 \pm 0.37$ \\
w/o Global Role Memory     & $90.72 \pm 0.51$ \\
w/o Node Memory            & $82.36 \pm 0.83$ \\
w/o Attention              & $87.91 \pm 0.64$ \\
w/o Team Assignment Head   & $89.43 \pm 0.56$ \\
\hline
\end{tabular}
\end{table}

\subsubsection{Computational Complexity Analysis} 
In a conventional TGN, each node \(i\) maintains a
\(d\)-dimensional temporal memory vector
\(\mathbf{m}_i(t) \in \mathbb{R}^d\), which summarizes its interaction history
up to time \(t\).
Given a batch containing \(N\) nodes and \(E\) temporal interactions, the dominant computational cost in TGN arises during the
neighbor aggregation step used for embedding computation,
where multi-head temporal attention applies learnable linear
transformations to node and edge representations. This
operation incurs a complexity of \(O(E d^2)\).
The memory update for each node, implemented using GRU, incurs an additional cost of \(O(N d^2)\).
Therefore, the overall computational complexity of the baseline TGN per batch is $O(E d^2 + N d^2)$, 
while the memory requirement for storing node states scales as \(O(N d)\).

The proposed MR-TGN augments the baseline TGN by introducing a small set of
\(K\) meta-role prototypes
\(\mathcal{R} = \{\mathbf{r}_1, \mathbf{r}_2, \ldots, \mathbf{r}_K\}\),
where each prototype \(\mathbf{r}_k \in \mathbb{R}^d\).
Each node computes relevance-weighted interactions with these role prototypes to obtain role-aware contextual information, resulting in an additional role-attention cost of
\(O(N K d)\).
The storage requirement for the role prototypes is \(O(K  d)\).
Since the number of roles \(K\) is small and independent of the
number of nodes \(N\) and interactions \(E\), the added computational and memory
overhead scales linearly with \(N\) and is negligible compared to the
quadratic terms in the baseline TGN complexity.
As a result, the proposed MR-TGN incurs
minimal additional cost.

\section{Conclusion}
This work addresses the problem of collective intent prediction in dynamic multi-agent environments, where heterogeneous agents exhibit coordinated behaviours, evolving interaction patterns, and only partially observable trajectories. To address these challenges, a MR-TGN framework was proposed for learning team-aware representations and predicting collective intent from dynamic interaction graphs. The framework integrates rich agent-level and pairwise interaction features with temporal graph representation learning, persistent memory, and globally shared role representations to capture instantaneous coordination patterns, historical dependencies, and recurring functional behaviours across heterogeneous agents. A sequential multi-task learning strategy was further employed to first discover the underlying team structure and subsequently exploit the learned team information for collective intent prediction.

The effectiveness of the proposed framework was evaluated using multi-agent scenarios involving heterogeneous missions and coordinated team behaviours under both full and partial observation settings. Comparative experiments with spatial, temporal, spatial--temporal, and dynamic graph-based models demonstrated the effectiveness of combining persistent temporal memory, interaction-aware representation learning, role information, and team-aware reasoning for collective intent prediction. In addition to conventional classification performance, the temporal behaviour of the models was evaluated using Early Correct Sample and Time To Stable Sample metrics, providing further insight into how early and reliably the underlying collective intent can be inferred from evolving observations. The experimental results showed that MR-TGN provides more accurate and temporally reliable predictions, particularly under scenarios where only partial observations of evolving multi-agent behaviors are available.

The analysis of the learned role representations further demonstrates that a compact set of globally shared role vectors can capture recurring functional behaviour patterns across different teams and mission contexts. The ablation studies confirm the contribution of the major architectural components and highlighted the importance of persistent memory, temporal attention, role-aware representation learning, and team-aware intent reasoning. These findings indicate that explicitly modelling both agent-level functional behaviours and dynamically emerging team structures provides an effective representation for reasoning about coordinated heterogeneous multi-agent systems.

Although air combat is used as a case study in this work, the proposed framework is applicable to a broad class of dynamic multi-agent systems characterized by heterogeneous agents, evolving coordination patterns, and incomplete observations. Future work will investigate online adaptation to previously unseen coordination strategies and intents, as well as uncertainty-aware prediction under noisy and incomplete observations. In addition, extending the framework towards zero-shot collective intent prediction and continual adaptation of role representations will be explored to further enhance its generalization capability in evolving multi-agent environments with novel coordination behaviors.



\bibliographystyle{cas-model2-names}

\bibliography{cas-refs}



\end{document}